\documentclass[12pt]{article}
\usepackage{geometry}
\usepackage{array}
\usepackage{longtable}
\usepackage{afterpage}
\usepackage{pdflscape}
\usepackage{arydshln}
\usepackage{multirow}
\usepackage{float}
\usepackage{url}
\usepackage{amsmath}
\usepackage{amsthm}
\usepackage{eufrak}
\usepackage{bbm}
\usepackage{graphicx}
\usepackage{esint}
\usepackage{setspace}
\usepackage{titlesec}
\usepackage{color}
\usepackage[acronym,nonumberlist]{glossaries}
\usepackage{natbib}

\titleformat{\section}{\large\bfseries}{\thesection}{1em}{}
\titleformat{\subsection}{\bfseries}{\thesubsection}{1em}{}
\titleformat{\subsubsection}{\itshape}{\thesubsubsection}{1em}{}

\theoremstyle{definition}

\newacronym{ap}{AP}{aggregation property}
\newacronym{ap1}{AP1}{aggregation property}
\newacronym{ap2}{AP2}{aggregation property}
\newacronym{di}{DI}{discretisation-invariant}
\newacronym{mtm}{MtM}{mark-to-market}
\newacronym{lv}{LV}{log variance}
\newacronym{otm}{OTM}{out-of-the-money}
\newacronym{pnl}{P\&L}{profit and loss}
\newacronym{qv}{QV}{quadratic variation}
\newacronym{rv}{RV}{realised variance}
\newacronym{snp}{S\&P 500}{Standard \& Poor's 500 Stock Market Index}
\newacronym{sdi}{SDI}{strike-discretisation-invariant}
\newacronym{vix}{VIX}{CBOE Volatility Index}

\begin{document}

\title{\Large Model-Free Discretisation-Invariant Swap Contracts}
\author{\large Carol Alexander and Johannes Rauch\footnote{School of Business, Management and Economics, University of Sussex, United Kingdom. Carol Alexander: c.alexander@sussex.ac.uk; Johannes Rauch: j.rauch@sussex.ac.uk.}}
\date{This version: April 2016}
\maketitle

\doublespacing

\thispagestyle{empty}

\begin{abstract}\normalsize
\noindent Realised pay-offs for discretisation-invariant swaps are those which satisfy a restricted  `aggregation property' of \cite{N12} for twice continuously differentiable deterministic functions of a multivariate martingale. They are initially characterised as solutions to a second-order system of PDEs, then those pay-offs based on martingale and log-martingale processes alone form a vector space. Hence there exist an infinite variety of other variance and higher-moment risk premia that are less prone to bias than standard variance swaps because their option replication portfolios have no discrete-monitoring or jump errors. Their fair values are also independent of the monitoring partition.  
A sub-class consists of pay-offs with fair values that are further free from numerical integration errors over option strikes. Here exact pricing and hedging is possible via dynamic trading strategies on a few vanilla puts and calls. An S\&P 500 empirical study on higher-moment and other DI swaps concludes.\\
\end{abstract}

\newpage


\clearpage
\glsresetall
\setcounter{page}{1}

\noindent 
Variance and volatility swaps, futures and options are popular instruments for diversifying investment portfolios and transferring volatility risk.\footnote{Variance swaps were introduced over-the-counter in the 1990's \citep{DDKZ99} and their futures, options, notes, funds and other derivatives are now being actively traded on exchanges, demand stemming from their role as a diversifier, a hedge or purely for speculation, as illustrated by \cite{AKK15}.} For instance, the terms and conditions of a conventional variance swap define the floating leg (realised variance) as the average squared daily log-return on some underlying, commonly an equity index, over the life of the swap. It is common practice for issuers to use the formula underlying the \gls{vix} for determining their swap rate,\footnote{Currently, CBOE data show that \$3-\$6bn notional is traded daily on VIX futures contracts alone and on stock exchanges around the world even small investors can buy and sell over a hundred listed products linked to volatility futures. The most popular of these is Barclay's VXX note, with a market cap of around \$1 trillion as of 31 December 2013.}
but this way the theoretical fair-value variance swap rate can only be approximated. Consequently, market rates can deviate well beyond the no-arbitrage range, especially during crisis periods, which is when trading in volatility products increases.\footnote{For example, during the financial crisis in 2008, market variance swap rates for the \gls{snp} were very often 5\% or more above the \gls{vix} -- see \cite{AK12} and \cite{KS14}.} These deviations can be attributed to a variety of discretisation and model-dependent errors, whose common effect is that theoretical prices for variance swaps can be unfair or even misleading. 

Sound theoretical prices for derivative contracts with complex pay-offs are important, because they help to preclude arbitrage opportunities, so there is a large and growing literature on approximation errors in variance swap rates, reviewed later.  
Taking an entirely different approach both \cite{N12} and \cite{B14} re-define the realised variance in such a way that there exists an exact, model-free fair-value variance swap rate under the minimal assumption of no arbitrage. Furthermore, \cite{N12} proves that this same rate applies irrespective of the monitoring frequency of the floating leg, 
provided his `aggregation property' (AP) holds for the pay-off. He defines one realised third moment for which the AP holds, and 
 an exact fair-value third moment swap rate exists which is independent of the monitoring frequency of the floating leg. The same applies to the new realised variance definitions in \cite{N12} and \cite{B14}.\footnote{He concludes by stating that ``[...] it would also be nice to be able to extend the analysis to higher-order moments. This would not be straightforward; [...] the set of functions that possess the aggregation property is quite limited; the way forward here may be to include other traded claims, in addition to those on the variance of the distribution."}

Pursuing these ideas we restrict the AP to twice continuously-differentiable pay-offs on adapted processes that contain only deterministic functions of martingale forward prices, thereby defining the class of  \gls{di} swap contracts. This way we can provide a comprehensive theory for DI swaps, written on multiple assets, which have exact fair-values, independent of the monitoring partition, provided only that the market is free of arbitrage opportunities.\footnote{They are `exact' in that they have no jump or discretisation biases, and so market swap rates should remain within the no-arbitrage range, even in times of financial distress, which is when the errors in standard variance swaps rates are considerable.} Our theory encompasses a wide variety of \gls{di} pay-offs, including those corresponding to higher moments of the log return distribution and bi-linear functions of vanilla options prices. We also describe dynamic trading strategies in a small number of vanilla-style contingent claims that allow one to hedge \gls{di} swaps in a model-free manner, and our empirical study applies these strategies to the \gls{snp}. 

In the following: Section \ref{sec:literature} sets our work in the context of the relevant literature and defines our notation; Section \ref{sec:diswaps} presents our theoretical results and describes the pricing and hedging of \gls{di} pay-offs; Section \ref{sec:empirical} presents the empirical results; Section \ref{sec:conc} concludes. Main proofs are in the Appendix.

\section{Background}\label{sec:literature}\glsresetall

A conventional variance swap of maturity $T$ defines the \gls{rv} as the average squared daily log return on some underlying over the term of the swap: 
\begin{equation}\label{eq:realisedvariance}
\mbox{RV}:=\sum_{t=1}^T\left(x_t-x_{t-1}\right)^2,
\end{equation}
where $x_t:=\ln F_t$ and $F_t>0$ denotes the underlying forward price at time $t$.\footnote{In practice, the floating leg of a variance swap is set equal to the \textit{average} realised variance taken over all trading days during the lifespan of the swap rather than the total variance as in \eqref{eq:realisedvariance}. However, including this level of detail would only add an unnecessary level of complexity to our analysis.} 
The calculation of a fair-value variance swap rate proceeds under the assumptions that the pricing measure is unique,\footnote{In an arbitrage-free market, as in \cite{HK79}, expected pay-offs may be computed in a risk-neutral measure. In a complete market the risk-neutral measure for a representative investor corresponds to a unique market implied measure, see \cite{BL78}.} and: (a) monitoring of the floating leg happens continuously; (b) the forward price of the underlying follows a pure diffusion process; (c)  vanilla options on the underlying with the same maturity as the swap are traded at a continuum of strikes. Then a unique and exact fair-value swap rate -- which under assumption (a) becomes the expected quadratic variation of the log price -- is derived from market prices of these  options. 

However, in the real world none of these assumptions hold. \cite{CW09} discuss the idealised case (a) where continuous monitoring is possible, replacing \eqref{eq:realisedvariance} by the \gls{qv} of log returns. Then 
they apply the replication theorem of \cite{CM01} to prove that, for a generic jump-diffusion process:
\begin{equation*}\label{eq:swapratejump}
\mathbbm{E}\left[\mbox{QV}\right]=2\int_{\mathbbm{R}^+}k^{-2}{q}(k)dk+\iota,
\end{equation*}
where $\mathbbm{E}$ denotes the expectation under the pricing measure and ${q}(k)$ denotes the price of a vanilla \gls{otm} option with strike $k$ and maturity $T$.\footnote{When $k\le F_{_0}$ the option is a put and when $k>F_{_0}$ the option is a call. This choice of separation strike is standard in the variance swap literature, e.g. in \cite{BKM03}.} When the underlying price follows a pure diffusion as in (b) the jump error $\iota$ is zero. Regarding assumption (c), in practice the integral in \eqref{eq:swapratediscrete} must be computed numerically, using the prices of vanilla options that are actually traded. \cite{JT05} address the problems attendant to this assumption and derive upper bounds for the so-called `truncation error'. Also based on a finite number of traded strikes, \cite{DO14} derive model-free arbitrage bounds for continuously-monitored variance swap rates and claim that market rates are surprisingly close to the lower bound. 

A major source of error in the fair-value swap rate stems from assumption (a) because floating legs must be monitored in discrete time. This `discrete-monitoring' error may be written
\begin{equation}\label{eq:discretemonitoringerror}
\delta:=\mathbbm{E}\left[\mbox{RV}-\mbox{QV}\right].
\end{equation}
Then, in the generic jump-diffusion setting of \cite{CW09}, the fair-value swap rate for the realised variance \eqref{eq:realisedvariance} may  be written
\begin{equation}\label{eq:swapratediscrete}
\mathbbm{E}\left[\mbox{RV}\right]=2\int_{\mathbbm{R}^+}k^{-2}{q}(k)dk+\iota+\delta.
\end{equation}
There is a large body of research on these pricing errors:  \cite{CL09} prove that the discrete monitoring error $\delta$ is related to the third moment of returns; \cite{JKLP13} investigate the convergence of the discretely-monitored swap rate to its continuously-monitored counterpart and derive bounds on $\delta$ that get tighter as the monitoring frequency increases; \cite{BCM14} generalise these results and provide conditions for signing $\delta$;  \cite{HK12} derive model-free  bounds for $\delta$; \cite{BJ08b} derive fair-value swap rates for discretely-monitored variance swaps under various stochastic volatility diffusion and jump models, claiming that for most realistic contract specifications $\delta$ is smaller than the error due to violation of assumption (b); \cite{BC14} extend their analysis to include a much wider variety of processes by considering the asymptotic expansion of $\delta$. Finally, \cite{RT13} derive bounds for the jump error $\iota$ and demonstrate, via simulations and an empirical study, that price jumps induce a systematic negative bias which is particularly apparent when there are large downward jumps.

\cite{N12} finds a way to  avoid the errors arising from assumptions (a) and (b): by discarding the conventional definition of realised variance  and using instead the log variance pay-off function $\lambda\left(\hat{x}\right):=2\left(\mathrm e^{\hat{x}}-1-\hat{x}\right)$ where $\hat{x}$ denotes the log return.\footnote{Note that the \gls{lv} can also be written as a function of the starting value $F$ and terminal value $F+\hat{F}$ of an increment in the underlying forward price, namely $\lambda^*\left(F,F+\hat{F}\right):=2\left[\tfrac{\hat{F}}{F}-\ln\left(\tfrac{F+\hat{F}}{F}\right)\right]$, where clearly $\lambda^*\left(F,F+\hat{F}\right)=\lambda\left(\hat{x}\right)$. Taylor expansion shows that the \gls{lv} may be associated with the second moment of the distribution of $\hat{x}$, since $\lim_{\hat{x}\rightarrow0}\lambda\left(\hat{x}\right)/\hat{x}^2=1$.} 
  The floating leg of Neuberger's log variance swap is defined as:\footnote{Other authors explore different definitions for the realised variance which give fair values that are easier to price and hedge than standard variance swap rates. \cite{M13} advocates the use of a sum of squared `simple' returns, rather than log returns, arguing that with this modification both jump and discretisation errors are minimised. Likewise, the gamma swaps described by \cite{L10} weight the realised variance in such a way that  replication and valuation are relatively straightforward under the continuous semi-martingale assumption. \cite{B14} derives generalised variance pay-offs that are also based on weighting functions. A common feature of these approaches is that they all re-define the floating leg based only on information about the underlying price.}
\begin{equation*}\label{eq:logvariance}
\mbox{LV}:=\sum_{t=1}^T\lambda \left(\hat{x}_t \right) = \sum_{t=1}^T\lambda\left(x_t-x_{t-1}\right).
\end{equation*}
With this definition, and under the minimal assumption that $F=\mathrm e^x$ follows a martingale under the risk-neutral measure (i.e. the market is free of arbitrage), the fair-value swap rate is free from both jump and discrete-monitoring errors. It is given by
\begin{equation*}\label{eq:logvarianceswaprate}
\mathbbm{E}\left[\mbox{LV}\right]=2\int_{\mathbbm{R}^+}k^{-2}{q}(k)dk.
\end{equation*}
The expected \gls{pnl} under the risk-neutral measure from investing in this variance swap is zero, and the same swap rate applies for all monitoring frequencies. In fact, the monitoring partition $\boldsymbol\Pi_{_N}$ used to determine the realised log variance
does not even have to be regular since
\begin{equation}\label{eq:logvarianceaggregation}
\mathbbm{E}\left[\sum_{\boldsymbol\Pi_{_N}}\lambda\left(\hat{x}\right)\right]=\mathbbm{E}\left[\lambda\left(x_{_T}-x_{_0}\right)\right] \quad \forall \,\, \boldsymbol\Pi_{_N},
\end{equation}
where $\boldsymbol\Pi_{_N}=\left\{0=t_{_0}<t_{_1}<\ldots<t_{_N}=T\right\}$ is a partition of the interval $\boldsymbol\Pi:=[0,T]$. From henceforth we write $A:=\left\{A_t\right\}_{t\in\boldsymbol\Pi}$ to denote the univariate process $A$ monitored over $\boldsymbol{\Pi}$, and for a multivariate process we write $\mathbf  {A}:=\left\{\mathbf {A}_t\right\}_{t\in\boldsymbol\Pi}$. Also $\mathbbm{E}_t[.]:=\mathbbm{E}[.|\mathcal{F}_t]$ denotes the  expectation conditional on the filtration at time $t$, with $\mathbbm{E}[.]:=\mathbbm{E}_0[.]$.


\cite{N12} introduced his `aggregation property' (AP) as follows:\footnote{\cite{N12} considers the case when the measure for \eqref{eq:aggregationproperty} is the pricing measure. See \cite{N12}, p.7: ``If the measure is a pricing measure, it says that the fair price of a one-month variance swap computed daily (a swap that pays the realized daily variance over a month) is the same as the price of a contingent claim that pays $\left(S_{_T}-S_{_0}\right)^2$. Indeed, because the relationship holds under any pricing measure (because the process is a martingale under any pricing measure), it also implies that a variance swap can be perfectly replicated if the contingent claim exists (or can be synthesised from other contingent claims) and the underlying asset is traded."} given $\phi:\mathbbm{R}^n\rightarrow\mathbbm{R}$ and an adapted process $\mathbf{z}\in\mathbbm{R}^n$, the pair $\left(\phi,\mathbf{z}\right)$ satisfies the \gls{ap} if and only if:\footnote{A simple lemma in Appendix B shows that \eqref{eq:aggregationproperty} is necessary for the absence of a discrete monitoring error.}
\begin{equation}\label{eq:aggregationproperty}
\mathbbm{E}\left[\sum_{\boldsymbol\Pi_{_N}}\phi\left(\mathbf{\hat{z}}\right)\right]=\mathbbm{E}\left[\phi\left(\mathbf{z}_{_T}-\mathbf{z}_{_0}\right)\right] \quad \forall \,\, \boldsymbol\Pi_{_N}.
\end{equation}

Two trivial cases are: (a) if $\phi$ is linear, say $\phi(\mathbf{\hat{z}})=\boldsymbol\alpha^\prime\mathbf{\hat{z}}$ for some $\boldsymbol\alpha\in\mathbbm{R}^n$, then \eqref{eq:aggregationproperty} holds for any process $\mathbf{z}$ because $\sum_{\boldsymbol\Pi_{_N}}\mathbf{\hat{z}}=\mathbf{z}_{_T}-\mathbf{z}_{_0}$; (b) if $\mathbf{z}$ contains only constant processes then $\hat{\mathbf{z}}_i=\mathbf{0}$ $\forall i\in\{1,\ldots,N\}$, so \eqref{eq:aggregationproperty} holds for any function with $\phi(\mathbf{0})=0$. Note that \eqref{eq:nodiscretemonitoringerror} also holds in case (a) because $\langle\mathbf{z}\rangle_{_T}^\phi=\mathbf{z}_{_T}-\mathbf{z}_{_0}$ and in case (b) because $\langle\mathbf{z}\rangle_{_T}^\phi=0$, provided $\phi(\mathbf{0})=0$.

The analogy between  \eqref{eq:aggregationproperty} and \eqref{eq:logvarianceaggregation} is obvious, and it is easy to see that the \gls{ap} does not hold for $\phi\left(\hat{x}\right)=\hat{x}^2$, the conventional variance pay-off.\footnote{In fact, the \gls{ap} does not hold for any $\phi\left(\hat{x}\right)=\hat{x}^n$, $n\ge 2$.
} Yet, if the \gls{ap} does hold, the r.h.s. of \eqref{eq:aggregationproperty} indicates that the expectation of the floating leg is path-independent, and even if investors differ in their views about jump risk in an incomplete market they will still agree on the fair-value swap rate. Furthermore, if the components of $\mathbf{z}$ only depend on the distribution of a single underlying asset with forward price process $F$, the fair-value swap rate can be expressed in terms of vanilla \gls{otm} options written on this asset by applying the replication theorem of \cite{CM01}. 

An alternative definition to \eqref{eq:aggregationproperty} of the AP is given in \cite{B14} and a simple characterisation of the process for which the two definitions are equivalent is presented in Lemma 1 of the Appendix. Interestingly, our analytic results on Theorems 2 and 3 below also require the same restricted characterisation, i.e. that the adapted process is given by $\mathbf{z}=\left(\mathbf{F},\mathbf{x}\right)^\prime$, where $\mathbf{x}:=\ln\mathbf{F}$ and $\mathbf{F}>\mathbf{0}$ denotes a vector of martingale forward prices.
While \cite{B14} pursues the univariate case, 
\cite{N12} takes the original step of including conditional fair-value processes of vanilla-style contingent claims in $\mathbf{z}$, allowing the floating leg of a swap to encompass information about serial dependence. He then considers all pay-off functions $\varphi$ which satisfy \eqref{eq:aggregationproperty} for 
$\mathbf{z}=\left(x,v\right)^\prime$, where  $x_t:=\ln F_t$, and $v$ denotes a generalised variance process $v_t:=\mathbbm{E}_t\left[\sigma\left(x_{_T}-x_t\right)\right]$ with $\sigma:\mathbbm{R}\rightarrow\mathbbm{R}$ and $\lim_{\hat{x}\rightarrow0}\sigma\left(\hat{x}\right)/\hat{x}^2=1$:
\begin{equation*}
\mathbbm{G}:=\left\{\varphi:\mathbbm{R}^{2}\rightarrow\mathbbm{R}\left|\varphi\left(\mathbf{\hat{z}}\right)=h_1\hat{x}+h_2\left(\mathrm{e}^{\hat{x}}-1\right)+h_3\hat{v}+h_4\left(\hat{v}-2\hat{x}\right)^2+h_5\left(\hat{v}+2\hat{x}\right)\mathrm{e}^{\hat{x}}\right.\right\},
\end{equation*}
subject to the restrictions $\sigma=\lambda$ if $h_4\ne 0$ and $\sigma=\eta$ if $h_5\ne 0$, where $\eta\left(\hat{x}\right):=2\left(\hat{x}\mathrm e^{\hat{x}}-\mathrm e^{\hat{x}}+1\right)$ denotes the `entropy variance'. The \gls{lv} pay-off relates to $h_1=-2$, $h_2=2$, $h_3=h_4=h_5=0$. Within the set $\mathbbm{V}$ of pay-off functions Neuberger further identifies the pay-off
\begin{equation*}
\psi\left(\mathbf{\hat{z}}\right):=
3\hat{v}\left(\mathrm e^{\hat{x}}-1\right)+\tau\left(\hat{x}\right),
\end{equation*}
with $\tau\left(\hat{x}\right):=6\left(\hat{x}\mathrm e^{\hat{x}}-2\mathrm e^{\hat{x}}+\hat{x}+2\right)$, which corresponds to $h_1=6$, $h_2=-12$, $h_3=-3$, $h_4=0$ and $h_5=3$, and argues that it approximates the third moment of log returns since $\lim_{\hat{x}\rightarrow 0}\tau\left(\hat{x}\right)/\hat{x}^3=1$. However, the first term does not vanish under expectation for partial increments even if $F$ follows a martingale. In fact it measures the covariance between returns and changes in implied variance. For the fair-value swap rate we have
\begin{equation*}
\mathbbm{E}\left[\psi\left(\mathbf{z}_{_T}-\mathbf{z}_{_0}\right)\right]=
\mathbbm{E}\left[\tau\left(x_{_T}-x_{_0}\right)\right],
\end{equation*}
which is dominated by the higher-order terms of $\tau$ for sufficiently large $x_{_T}-x_{_0}$. Therefore the association of either the floating or the fixed leg of this swap with the third moment is questionable.\footnote{c.f. p.3435 in \cite{N12}, Proof of Proposition 6.}
The subsequent empirical study of \cite{KNS13} shows that the \gls{pnl} on the skewness swap based on $\mathbbm{G}$ is strongly correlated with that on a variance swap. The flexibility to define a great variety of swap contracts with potentially diverse \glspl{pnl} and model-free swap rates that are independent of the monitoring frequency motivates our research.

\section{Discretisation-Invariant Swap Contracts}\label{sec:diswaps}\glsresetall

By restricting the definition of the AP in \cite{N12} to  $\phi\in\mathcal{C}^2$ with $\phi(\mathbf{0})=0$, and additionally to a multivariate stochastic process $\mathbf{z}\in\mathbbm{R}^n$ containing only deterministic functions of martingale forward prices $\mathbf{F}  \in\mathbbm{R}^d$ of $d$ tradable assets or derivatives in an arbitrage-free market,\footnote{For instance, the process $\mathbf{z}$ may contain futures prices and/or the logs of these prices. We make the minimal no-arbitrage assumption only to ensure that futures prices follow a multivariate $\mathbbm{Q}$-martingale.} we can characterise all `discretisation-invariant' swap contracts as solutions to a multivariate second-order PDE system. With the further restriction that $\mathbf{z}=\left(\mathbf{F}, \mathbf{x}\right)^{\prime}$  there exists an entire vector space of DI swaps with analytic pay-offs $\phi\left(\mathbf{\hat{z}}\right)$. Interestingly, this same restriction  also unifies the AP of \cite{B14} with that of \cite{N12} as shown in the Appendix. 

These DI swaps may give access to a great variety of risk premia, including premia associated with more complex trading strategies than simple moments. In particular, rather than a single definition for realised skewness as in \cite{N12}, we  obtain infinitely many pay-offs with aggregating characteristics, and which may  therefore be exactly priced.

The term `swap' here is used in a generic sense, as follows: given a pay-off $\phi:\mathbbm{R}^n\rightarrow\mathbbm{R}$ and $\mathbf{z}$, the floating leg of a `$\phi$-swap' w.r.t. a partition $\boldsymbol\Pi_{_N}$ is defined as\footnote{\cite{N12} calls the pay-off a `characteristic' while \cite{B14} simply refers to a `function'.}
\begin{equation}\label{eq:floatingleg}
\sum_{\boldsymbol\Pi_{_N}}\phi\left(\mathbf{\hat{z}}\right):=\sum_{i=1}^N\phi\left(\mathbf{z}_{t_i}-\mathbf{z}_{t_{i-1}}\right).
\end{equation}
We consider only one maturity date, $T$, but various partitions of $\boldsymbol\Pi$, the standard one being the `daily' partition $\boldsymbol\Pi_{_D}:=\left\{0,1,\ldots,T\right\}$. 
The increments along a partition are denoted using a `carat'. 
Let $\left\{\boldsymbol\Pi_{_N}\right\}_{N=1,2,\ldots}$ denote a sequence of partitions such that $0=t_{_0}<t_{_1}<\ldots<t_{_N}=T$. If    $\max_{i\in\{1,\ldots,N\}}\left[t_i-t_{i-1}\right]\rightarrow 0$ as $N\rightarrow\infty$ we write $\boldsymbol\Pi_{_N}\rightarrow\boldsymbol\Pi$. If it exists we define the `$\phi$-variation' of $\mathbf{z}$ as the continuously monitored limit of the realised leg, i.e.
\begin{equation}\label{eq:fvariation}
\langle\mathbf{z}\rangle_{_T}^\phi:=\lim_{\boldsymbol\Pi_{_N}\rightarrow\boldsymbol\Pi}\sum_{\boldsymbol\Pi_{_N}}\phi\left(\mathbf{\hat{z}}\right).
\end{equation}
Since $\phi(\mathbf{0})=0$ a finite limit  \eqref{eq:fvariation} can exist, but we do not need to assume this because it does not preclude the definition of a `$\phi$-swap' as a financial contract that exchanges the realised leg \eqref{eq:floatingleg} with a fixed swap rate'.\footnote{The $\phi$-variation is a theoretical construct that, if it exists, can be used to derive a fair-value swap rate by taking its expected value based on some assumed  process for the underlying. This is the approach taken by \cite{JKLP13} and several other papers that analyse the discrete monitoring error for variance swaps.} However, if the $\phi$-variation exists and is finite the discrete monitoring error for a $\phi$-swap under the partition $\boldsymbol\Pi_{_N}$ may be written
\begin{equation}\label{eq:fdiscretemonitoringerror}
\delta_{_N}(\phi, \mathbf{z}):=\mathbbm{E}\left[\sum_{\boldsymbol\Pi_{_N}}\phi\left(\mathbf{\hat{z}}\right)-\langle\mathbf{z}\rangle_{_T}^\phi\right].
\end{equation}
Note that with $\mathbf{z}=x$ and $\phi(\hat{x})=\hat{x}^2$ the definition \eqref{eq:fvariation} corresponds to the QV of the log price and the discrete monitoring error is given by \eqref{eq:discretemonitoringerror}. 
Our focus is on those combinations $\left(\phi,\mathbf{z}\right)$ for which the discrete monitoring error $\delta_{_N}(\phi, \mathbf{z})$ is zero, i.e.
\begin{equation}\label{eq:nodiscretemonitoringerror}
\mathbbm{E}\left[\sum_{\boldsymbol\Pi_{_N}}\phi\left(\mathbf{\hat{z}}\right)\right]=\mathbbm{E}\left[\langle\mathbf{z}\rangle_{_T}^\phi\right]\quad \forall \,\, \boldsymbol\Pi_{_N}.
\end{equation}

\subsection{Characterisation of DI Swaps}

Let $\boldsymbol\Delta\in\mathbbm{R}^{n\times d}$ and $\boldsymbol\Gamma\in\mathbbm{R}^{n\times d\times d}$ denote the first and second partial derivatives of $\mathbf{z}$ w.r.t. $\mathbf{F}$ and denote by  $\mathbf{J}\left(\mathbf{\hat{z}}\right)\in\mathbbm{R}^n$ the Jacobian vector and $\mathbf{H}\left(\mathbf{\hat{z}}\right)\in\mathbbm{R}^{n\times n}$ the Hessian matrix of first and second partial derivatives of $\phi$ w.r.t. $\mathbf{\hat{z}}$. Our first result gives a joint condition on  $\phi$ and  the underlying dynamics $\mathbf{z}$ for the \gls{ap} to hold. Specifically, we derive a second order system of partial differential equations that represents a necessary condition, which is also sufficient for $\left(\phi,\mathbf{z}\right)$ to define a \gls{di} swap when $\mathbf{z}$ is a multivariate diffusion with finite $\phi$-variation.\vspace{12pt}

\noindent {\bf Theorem 1:} 
If $\left(\phi,\mathbf{z}\right)$ is such that either \eqref{eq:aggregationproperty} is true, or the $\phi$-variation of $\mathbf{z}$ exists and \eqref{eq:nodiscretemonitoringerror} is true, then the following second-order system of partial differential equations holds:
\begin{equation}\label{eq:equivalence}
\left[\mathbf{J}\left(\mathbf{\hat{z}}\right)-\mathbf{J}\left(\mathbf{0}\right)\right]^\prime\boldsymbol\Gamma+\boldsymbol\Delta^\prime\left[\mathbf{H}\left(\mathbf{\hat{z}}\right)-\mathbf{H}\left(\mathbf{0}\right)\right]\boldsymbol\Delta=\mathbf{0}.
\end{equation}
Further, if $\mathbf{F}$ follows a diffusion with finite $\phi$-variation then \eqref{eq:nodiscretemonitoringerror}, \eqref{eq:aggregationproperty} and \eqref{eq:equivalence} are equivalent.\vspace{12pt}

\noindent For a given $\mathbf{z}$ the above system may be solved numerically to yield all available \gls{di} pay-off functions $\phi$. However, pay-offs defined in terms of numerical procedures are difficult to monitor; indeed in practice we are only interested in the real, analytic solutions of \eqref{eq:equivalence}. To this end we provide Theorem 2, which is proved in the Appendix by solving \eqref{eq:equivalence} for a particular $\mathbf{z}$ and then showing, by straightforward evaluation of \eqref{eq:nodiscretemonitoringerror}, that the necessary condition is sufficient. It defines a vector space $\mathbbm{F}$ of DI pay-off functions for general underlying variables $\mathbf{F}$. For instance, we can include the log contract $X_t:=\mathbbm{E}_t\left[x_{_T}\right]$, the entropy contract $Y_t:=\mathbbm{E}_t\left[F_{_T}x_{_T}\right]$ or the conditional fair-value process of any other contingent claim in $\mathbf{F}$. The components of $\mathbf{F}$ can depend on one or more underlying assets, and it is possible to define \gls{di} covariance swaps using pay-offs from $\mathbbm{F}$, as well as other swap contracts that depend on a multivariate distribution. \vspace{12pt}

\noindent {\bf Theorem 2:} 
Let $\mathbf{F}>\mathbf{0}$ follow a $d$-dimensional martingale process and set $\mathbf{z}=\left(\mathbf{F},\mathbf{x}\right)^\prime$ with $\mathbf{x}:=\ln\mathbf{F}$.\footnote{Here and in the following the vector notation $\ln\mathbf{F}$ as well as $\mathrm e^\mathbf{x}$ is understood component-wise.} Then the solutions to \eqref{eq:equivalence} form a vector space over $\mathbbm{R}$, defined by:\footnote{Note that $\text{tr}\left(\boldsymbol\Omega\mathbf{\hat{F}}\mathbf{\hat{F}}^\prime\right)$ may be written as the quadratic form $\mathbf{\hat{F}}^\prime\boldsymbol\Omega\mathbf{\hat{F}}$ so we may assume $\boldsymbol\Omega=\boldsymbol\Omega^\prime$ w.l.o.g..}
\begin{equation*}
\mathbbm{F}:=\left\{\phi:\mathbbm{R}^{n}\rightarrow\mathbbm{R}\left|\phi\left(\mathbf{\hat{z}}\right)=\boldsymbol\alpha^\prime\mathbf{\hat{F}}+\text{tr}\left(\boldsymbol\Omega\mathbf{\hat{F}}\mathbf{\hat{F}}^\prime\right)+\boldsymbol\beta^\prime\left(\mathrm e^{\mathbf{\hat{x}}}-\mathbf{1}\right)+\boldsymbol\gamma^\prime\mathbf{\hat{x}}\right.\right\},
\end{equation*}
where $\boldsymbol\alpha$, $\boldsymbol\beta$, $\boldsymbol\gamma\in\mathbbm{R}^{d}$ and $\boldsymbol\Omega=\boldsymbol\Omega^\prime\in\mathbbm{R}^{d\times d}$.\vspace{12pt}

\noindent Theorem 2 includes pay-offs that are linear and quadratic in the components of $\mathbf{F}$  and linear in the log and percentage returns, i.e. $\mathbf{\hat{x}}$ and $\mathrm e^\mathbf{\hat{x}}-\mathbf{1}$, respectively. Of course, we can include any martingale in $\mathbf{F}$ and later we shall use the fair-value processes of power log contracts to construct $\phi$-swaps with realised pay-offs that correspond to higher moments of log returns.\footnote{Note that with $\mathbf{F}=\left(F,X\right)^\prime$, we can relate the variance pay-off functions introduced by \cite{N12} to specific pay-offs in $\mathbbm{F}$. For instance, the \gls{lv} pay-off can be obtained by choosing $\boldsymbol\alpha=\mathbf{0}$, $\boldsymbol\Omega=\mathbf{0}$, $\boldsymbol\beta=\left(2,0\right)^\prime$, and $\boldsymbol\gamma=\left(-2,0\right)^\prime$.}

In a wider sense all self-financing portfolios are \gls{di} because their expected profit in an arbitrage-free market is zero, irrespective of the frequency of trading. It is possible to relax the assumption that $\phi\in\mathcal{C}^2$, so that  $\mathbbm{F}$ can include pay-offs $\boldsymbol\alpha\left(\mathbf{F}_{t-1}\right)^\prime\mathbf{\hat{F}}_t$ that are functions of both the increment and the starting value. These represent piecewise dynamic trading strategies in the components of $\mathbf{F}$.  For instance, percentage returns as well as quadratic pay-offs correspond to specific dynamic trading strategies. Also under these relaxed assumptions, the third moment pay-off from \cite{N12} would be included in $\mathbbm{F}$. Otherwise this pay-off provides an example of an AP characteristic which is not a DI pay-off.\footnote{It may be written as a dynamic trading strategy in $\mathbf{F}=\left(F,X,Y\right)^\prime$, where $X$ and $Y$ are the log and entropy contracts respectively, with $\boldsymbol\alpha\left(\mathbf{F}_{t-1}\right)=\left(-12F^{-1}_{t-1}-6F^{-2}_{t-1}Y_{t-1},6,6F^{-1}_{t-1}\right)^\prime$, $\boldsymbol\Omega=\mathbf{0}$ and $\boldsymbol\beta=\boldsymbol\gamma=\mathbf{0}$.} It is those pay-offs associated with $\boldsymbol\Omega$, which require the trading of contracts not included in $\mathbf{F}$, that we focus on in the following.

\subsection{Pricing and Hedging DI Swaps}

The fixed leg of a $\phi$-swap corresponds to the risk-neutral expectation of the floating leg at inception, and the fair-value swap rate for a \gls{di} swap is given by $v^\phi_{_0}:=\mathbbm{E}\left[\phi\left(\mathbf{z}_{_T}-\mathbf{z}_{_0}\right)\right]$. We now consider the conditional fair-value process 
$V^\phi_t:= \mathbbm{E}_t\left[\sum_{\boldsymbol\Pi_{_N}}\phi\left(\mathbf{\hat{z}}_i\right)\right]-v^\phi_{_0}$, from marking the \gls{pnl} to market, which is typically done at the end of each trading day. 
Note that the \gls{ap} implies $V^\phi_{_0}=0$, and that $V^\phi_{_T}$ is the total \gls{pnl} on the swap at maturity. From henceforth we use the daily partition $\boldsymbol\Pi_{_D}$ in the text, for ease of exposition,  while all proofs in the Appendix are for general $\boldsymbol\Pi_{_N}$. 

When hedging the swap we seek to replicate the increment $\hat{V}^\phi_t:=V^\phi_t-V^\phi_{t-1}$, for which the following is useful:\vspace{12pt} 

\noindent {\bf Theorem 3:} 
For $t\in\boldsymbol\Pi_{_D}$ the increments in the value process of a \gls{di} swap may be written
\begin{equation}\label{eq:realisedimplied}
\hat{V}^\phi_t=\phi\left(\mathbf{\hat{z}}_t\right)+\hat{v}^\phi_t,
\end{equation}
where  $v^\phi_t:=\mathbbm{E}_t\left[\phi\left(\mathbf{z}_{_T}-\mathbf{z}_t\right)\right]$ denotes the fair-value swap rate for the residual time-to-maturity. Further, when $\mathbf{z}=\left(\mathbf{F},\mathbf{x}\right)^\prime$ as in Theorem 2 we have
\begin{equation}\label{eq:theorem3}
\hat{V}^\phi_t=\boldsymbol\alpha^\prime\mathbf{\hat{F}}_t+\text{tr}\left(\boldsymbol\Omega\left[\boldsymbol{\hat{\Sigma}}_t-2\mathbf{F}_{t-1}\mathbf{\hat{F}}_t^\prime\right]\right)+\boldsymbol\beta^\prime\left(\mathrm e^{\mathbf{\hat{x}}_t}-\mathbf{1}\right)+\boldsymbol\gamma^\prime\mathbf{\hat{X}}_t,
\end{equation}
where 
$\boldsymbol{\Sigma}_t:=\mathbbm{E}_t\left[\mathbf{F}_{_T}\mathbf{F}_{_T}^\prime\right]$ and 
 $\mathbf{X}_t:=\mathbbm{E}_t\left[\mathbf{x}_{_T}\right]$.
The corresponding fair-value swap rate at inception is $v^\phi_{_0}=\text{tr}\left(\boldsymbol\Omega\left[\boldsymbol\Sigma_{_0}-\mathbf{F}_{_0}\mathbf{F}_{_0}^\prime\right]\right)+\boldsymbol\gamma^\prime\left(\mathbf{X}_{_0}-\mathbf{x}_{_0}\right)$.\vspace{12pt}

\noindent Theorem 3 characterises the \gls{pnl} which accrues to the issuer of a \gls{di} swap who pays fixed and receives floating. 
The decomposition \eqref{eq:realisedimplied} separates the change in the realised pay-off from the change in the implied leg. While the value process  follows a $\mathbbm{Q}$-martingale, the two components are generally not $\mathbbm{Q}$-martingales by definition.\footnote{Theorem 3 implies that, in order to represent an investable trading strategy, the conversion into constant maturity increments (as in our empirical study) has to be performed on the change in the swap value rather than the two components separately. For instance, in the case of Neuberger's variance swap the change in the swap value is the sum of the realised pay-off function $\lambda\left(\hat{x}\right)$ and the change in the swap rate $\hat{v}^\lambda$.} 
The swap can be hedged in discrete time using a static trading strategy in $\boldsymbol{\Sigma}$ and $\mathbf{X}$ and a dynamic trading strategy in $\mathbf{F}$, with dynamic hedging taking place along the monitoring partition $\boldsymbol\Pi_{_N}$. For instance, the \gls{pnl} on a swap based on the \gls{lv} is $\hat{V}^\lambda_t=2\left(\mathrm e^{\hat{x}_t}-1-\hat{X}_t\right)$ so, for $t\in\boldsymbol\Pi_{_N}$, 
$V^\lambda_t=2\sum_{i=1}^tF_{i-1}^{-1}\hat{F}_{i}-2\left(X_t-X_{_0}\right)$. 
Hence this swap can be hedged by buying two log contracts at initiation and dynamically rebalancing the position in the log contract, i.e. shorting $2F_{t-1}^{-1}$ futures contracts from time $t-1$ to $t$. 

The hedge specified by \eqref{eq:theorem3} contains static and dynamic delta elements. Since $\mathbf{\hat{F}}$ and $\mathbf{\hat{X}}$ correspond to price changes in portfolios that do not change over time, $\boldsymbol\alpha$ and $\boldsymbol\gamma$ are static hedge ratios. However, the holdings of the underlying which are determined variably by the previous prices $\mathbf{F}_{t-1}$ need to be dynamically rebalanced and hence $\boldsymbol\Omega$ and implicitly $\boldsymbol\beta$ are part of a dynamic hedge. These hedge ratios may change whenever the swap is monitored, and hedging is exact if rebalancing coincides with the monitoring partition of the swap. 

Pricing DI swaps is straightforward, given the following corollary, proved in the Appendix:\vspace{12pt}

\noindent {\bf Corollary:} 
The fair-value swap rate for a \gls{di} $\phi$-swap 
is
\begin{equation*}
v^\phi_{_0}=\text{tr}\left(\boldsymbol\Omega\left[\boldsymbol\Sigma_{_0}-\mathbf{F}_{_0}\mathbf{F}_{_0}^\prime\right]\right)+\boldsymbol\gamma^\prime\left(\mathbf{X}_{_0}-\mathbf{x}_{_0}\right).
\end{equation*}

\noindent Note that $v^\phi_{_0}$ is independent of $\boldsymbol\alpha$ and $\boldsymbol\beta$, since the corresponding pay-offs have zero expectation under the risk-neutral measure.

\vspace{12pt}

\noindent In the next section we shall consider  $n$-th power log contracts, i.e.   $X_t^{(n)}:=\mathbbm{E}_t\left[x_{_T}^n\right]$.\footnote{We assume they are tradable over-the-counter, but their replication portfolios are not exact, so transaction costs should be considered in practice.} According to the replication theorem of \cite{CM01}, this conditional expectation can be expressed in terms of vanilla \gls{otm} options as:
\begin{equation}\label{eq:powerlogcontracts}
X_t^{(n)}=x_t^n+\int_{\mathbbm{R}^+}\gamma_n(k)q_t(k)dk,
\end{equation}
where $\gamma_n(k):=n(\ln k)^{n-2}k^{-2}\left[n-1-\ln k\right]$ and $q_t(k)$ denotes the time-$t$  price of a vanilla \gls{otm} option with strike $k$ and maturity $T$. The following table shows replication portfolios for the first four power log contracts:\\
\begin{table}[h!]\small
	\begin{center}
		\begin{tabular}{l|lr}
			{\bf Contract} & {\bf Variable} & {\bf Pricing Formula}\\\hline
			\\
			Log & $X_t=$ & ${x_t}-\int_{\mathbbm{R}^+}k^{-2}q_t(k)dk $\\
			\\
			Squared log & $X_t^{(2)}=$ & ${x_t^2}+2\int_{\mathbbm{R}^+}\left(1-\ln k\right)k^{-2}q_t(k)dk$\\
			\\
			Cubed log & $X_t^{(3)}=$ & ${x_t^3}+3\int_{\mathbbm{R}^+}\ln k\left(2-\ln k\right)k^{-2}q_t(k)dk$\\
			\\
			Quartic log & $X_t^{(4)}=$ & ${x_t^4}+4\int_{\mathbbm{R}^+}(\ln k)^2\left(3-\ln k\right)k^{-2}q_t(k)dk$\\\hline
		\end{tabular}
	\end{center}
	{\caption[Fundamental contracts]{\small The first four power log contracts and their replication portfolios.}\label{tab:fundamentalcontractspricing}}
\end{table}

\noindent We may also consider the alternative replication scheme:
\begin{equation*}\label{eq:powerlogcontractsalt}
X_t^{(n)}=x_{_0}^n+nx_{_0}^{n-1}\left(\tfrac{F_t- F_{_0}}{F_{_0}}\right)+\int_0^{F_{_0}}\gamma_n(k)P_t(k)dk+\int_{F_{_0}}^\infty\gamma_n(k)C_t(k)dk,
\end{equation*}
where $P_t(k)$ and $C_t(k)$ denote the time-$t$ forward prices of vanilla put and call options with strike $k$ and maturity $T$. The difference between the two replication schemes is that \eqref{eq:powerlogcontracts} is based only on \gls{otm} options but due to the stochastic separation strike $F_t$ this portfolio would require continuous rebalancing between puts and calls. The alternative replication scheme involves options that are \gls{otm} only at inception and this portfolio describes buy-and-hold strategies that require no dynamic rebalancing. The two representations are exchangeable, and which is used  depends on the application. Most authors in this area employ  \cite{CM01} replication for pricing; the alternative may be preferable for static hedging.   

\subsection{Moment Swaps}\label{sec:ms}
For the next result we suppose that $\mathbf{F}$ contains power log contracts whose corresponding replication portfolios may be derived from \eqref{eq:powerlogcontracts}. 
Let $\mathbf{F}_t:=\left(X_t,X_t^{(2)}\ldots,X_t^{(n-1)}\right)^\prime$ for some $n\ge 2$ and consider the parameters

\begin{equation*}\label{eq:theorem4}
\boldsymbol\alpha=\boldsymbol\beta=\boldsymbol\gamma=\mathbf{0}\quad\text{and}\quad\boldsymbol\Omega=\boldsymbol\Omega^{(n)}:=\left[
\begin{array}{cccc}
\omega_1^{(n)}&\tfrac{1}{2}\omega_2^{(n)}&\ldots&\tfrac{1}{2}\omega_{n-1}^{(n)}\\
\tfrac{1}{2}\omega_2^{(n)}&0&\ldots&0\\
\vdots&\vdots&\ddots&\vdots\\
\tfrac{1}{2}\omega_{n-1}^{(n)}&0&\ldots&0\\
\end{array}\right],
\end{equation*}
with $\omega_{n-1}^{(n)}=1$ and
\begin{equation*}
\omega_i^{(n)}=X_{_0}^{n-1-i}\sum_{j=i+1}^n\tbinom{n}{j}\left(-1\right)^{n-j}=-X_{_0}^{n-1-i} \sum_{j=0}^i\tbinom{n}{j}(-1)^{n-j},
\end{equation*}
for $i\in\{1,\ldots,n-2\}$. Note that $\sum_{j=0}^n\tbinom{n}{j}(-1)^{n-j}=0$, so the swap capture the $n$-th (central) moment of the log-return distribution of $F$
\begin{equation*}
v^\phi_{_0}=\mathbbm{E}\left[\left(x_{_T}-X_{_0}\right)^n\right]=\sum_{i=1}^n\tbinom{n}{i}\left(-X_{_0}\right)^{n-i}X_{_0}^{(i)}+\left(-X_{_0}\right)^n :=v^{(n)}_{_0},
\end{equation*}
Using Theorem 3 we can derive the following hedging rule for \gls{di} moment swaps:
\begin{equation*}
\hat{V}_t^{(n)}:= \mathbbm{E}_t\left[\sum_{\boldsymbol\Pi_{_N}}\text{tr}\left(\boldsymbol\Omega^{(n)}\mathbf{\hat{F}}\mathbf{\hat{F}}^\prime\right)\right]-v^{(n)}_{_0}
=\sum_{i=1}^{n-1}\omega_i^{(n)}\left[\hat{X}_t^{(i+1)}-X_{t-1}\hat{X}_t^{(i)}-X_{t-1}^{(i)}\hat{X}_t\right],
\end{equation*}
where
\begin{equation*}
\boldsymbol\Omega^{(2)}=1,\quad
\boldsymbol\Omega^{(3)}=\left[
\begin{array}{cc}
-2X_{_0}&\tfrac{1}{2}\\
\tfrac{1}{2}&0
\end{array}\right],\quad
\boldsymbol\Omega^{(4)}=\left[
\begin{array}{ccc}
3X_{_0}^2&-\tfrac{3}{2}X_{_0}&\tfrac{1}{2}\\
-\tfrac{3}{2}X_{_0}&0&0\\
\tfrac{1}{2}&0&0
\end{array}\right],
\end{equation*}
 and we assume $\boldsymbol\alpha=\boldsymbol\beta=\boldsymbol\gamma=\mathbf{0}$ throughout.  
Then the realised characteristics for second, third and fourth moment DI  higher-moment swaps are reported in Table \ref{tab:momentswaps}, along with their fair-values, computed using the Corollary. For the hedging we suggest the dynamic trading strategies shown in Table \ref{tab:swaprates}, 
i.e. the variance swap can be hedged by selling a squared log contract and dynamically holding $2X_{t-1}$ log contracts, the third-moment swap can be hedged by selling a cubed log contract and dynamically holding $h_{2t}^{(3)}$ squared log contracts as well as $h_{1t}^{(3)}$ log contracts, and the fourth-moment swap can be hedged by selling a quartic log contract and holding $h_{3t}^{(4)}$ cubed log contracts, $h_{2t}^{(4)}$ squared log contracts and $h_{1t}^{(4)}$ log contracts from  $t-1$ to $t$.
\begin{table}[h!]\small
	\begin{center}
		\begin{tabular}{l|llr}
			{\bf Moment} & {\bf Parameters} & {\bf Floating Leg} & {\bf Fixed Leg}\\\hline
			\\
			Second & $\boldsymbol\Omega=\boldsymbol\Omega^{(2)}$ & $\sum_{\boldsymbol\Pi_{_N}}\hat{X}^2_i$ & $v^{(2)}_{_0}$\\
			\\
			Third & $\boldsymbol\Omega=\boldsymbol\Omega^{(3)}$ & $\sum_{\boldsymbol\Pi_{_N}}\left(\hat{X}^{(2)}_i\hat{X}_i-2X_{_0}\hat{X}^2_i\right)$ & $v^{(3)}_{_0}$\\
			\\
			Fourth & $\boldsymbol\Omega=\boldsymbol\Omega^{(4)}$ & $\sum_{\boldsymbol\Pi_{_N}}\left(\hat{X}^{(3)}_i\hat{X}_i-3X_{_0}\hat{X}^{(2)}_i\hat{X}_i+3X_{_0}^2\hat{X}^2_i\right)$ & $v^{(4)}_{_0}$\\\hline
		\end{tabular}
	\end{center}
	{\caption[Fundamental contracts]{\small Realised characteristics for DI moment swaps   with  fair values $v^{(2)}_{_0}=X_{_0}^{(2)}-X_{_0}^2$,
			 $v^{(3)}_{_0}=X_{_0}^{(3)}-3X_{_0}^{(2)}X_{_0}+2X_{_0}^3$ and $v^{(4)}_{_0}= X_{_0}^{(4)}-4X_{_0}^{(3)}X_{_0}+6X_{_0}^{(2)}X_{_0}^2-3X_{_0}^4$. }\label{tab:momentswaps}}
\end{table}

\begin{table}[h!]\small
	\begin{center}
		\begin{tabular}{l|lr}
			{\bf Moment} & {\bf Variable} & {\bf Hedging Strategy}\\\hline
			\\
			Second & $\hat{V}_t^{(2)}=$ & $\hat{X}_t^{(2)}-2X_{t-1}\hat{X}_t$\\
			\\
			Third & $\hat{V}_t^{(3)}=$ & $\hat{X}_t^{(3)}-h_{2t}^{(3)}\hat{X}_t^{(2)}-h_{1t}^{(3)}\hat{X}_t$\\
			\\
			Fourth & $\hat{V}_t^{(4)}=$ & $\hat{X}_t^{(4)}-h_{3t}^{(4)}\hat{X}_t^{(3)}-h_{2t}^{(4)}\hat{X}_t^{(2)}-h_{1t}^{(4)}\hat{X}_t$\\\hline
		\end{tabular}
	\end{center}
	{\caption[Fundamental contracts]{\small Trading strategies for the model-free hedging of DI moment swap contracts, where $h_{2t}^{(3)}:=2X_{_0}+X_{t-1}$, $h_{1t}^{(3)}:=X_{t-1}^{(2)}-4X_{_0}X_{t-1}$, $h_{3t}^{(4)}:=3X_{_0}+X_{t-1}$, $h_{2t}^{(4)}:=-3X_{_0}^2-3X_{_0}X_{t-1}$ and $h_{1t}^{(4)}:=X_{t-1}^{(3)}-3X_{_0}X_{t-1}^{(2)}+6X_{_0}^2X_{t-1}$.}\label{tab:swaprates}}
\end{table}

\subsection{Straddle Swaps}

All examples of \gls{di} swaps considered so far require integration over a continuum of strikes for valuing the fixed leg, but in practice options are traded for a relatively small number of discrete strikes. So this section introduces a class of DI swaps that can be priced and replicated exactly based only on the available options prices. 
Like all other \gls{di} swaps they have the same fair-value swap rate, independent of the monitoring partition $\boldsymbol{\Pi}_{_N}$, which is free from both discrete monitoring and model-specific (e.g. jump) errors. In addition, they do not rely on the replication of synthetic contingent claims such as power log contracts and hence there is no numerical integration error.

Let $\mathbf{F}=\left(\mathbf{P}, \mathbf{C}\right)^\prime$ where $\mathbf{P}:=\left\{\mathbf{P}_t\right\}_{t\in\boldsymbol\Pi}$ and $\mathbf{C}:=\left\{\mathbf{C}_t\right\}_{t\in\boldsymbol\Pi}$ describe the forward price processes of $d$ vanilla put options and $d$ vanilla call options, with identical, traded strikes $\mathbf{k}$, on an underlying futures with maturity $T$, so
 $
\mathbf{P}_t:=\mathbbm{E}_t\left[\left(\mathbf{k}-F_{_T}\mathbf{1}\right)^+\right]$ and $
\mathbf{C}_t:=\mathbbm{E}_t\left[\left(F_{_T}\mathbf{1}-\mathbf{k}\right)^+\right]
$ 
where $\mathbf{1}:=(1,\ldots,1)^\prime\in\mathbbm{R}^d$. Assume w.l.o.g. that the traded strikes $\mathbf{k}:=\left(k_1,\ldots,k_d\right)^\prime\in\mathbbm{R}^d$ are ordered such that $k_1<k_2<\ldots<k_d$, and denote by $\mathbf{\hat{P}}$ and $\mathbf{\hat{C}}$ the increments in $\mathbf{P}$ and $\mathbf{C}$, respectively. 
Let $\boldsymbol{\tilde\Omega}\in\mathbbm{R}^{d\times d}$ be a lower triangular matrix and set
\begin{equation*}
\boldsymbol\alpha = \boldsymbol\beta=\boldsymbol\gamma=\mathbf{0},\quad
\boldsymbol\Omega=\boldsymbol\Omega^S:=\left[
\begin{array}{cc}
\mathbf{0}&\tfrac{1}{2}\boldsymbol{\tilde\Omega}\\
\tfrac{1}{2}\boldsymbol{\tilde\Omega}^\prime&\mathbf{0}
\end{array}\right]\in\mathbbm{R}^{2d\times 2d}
\end{equation*}
Since the strikes are in ascending order either the put or the call has zero pay-off, so
\begin{equation*}
\mathbbm{E}\left[\text{tr}\left(\boldsymbol\Omega^S\mathbf{F}_{_T}\mathbf{F}_{_T}^\prime\right)\right]=\mathbbm{E}\left[\mathbf{P}_{_T}^\prime\boldsymbol{\tilde\Omega}\mathbf{C}_{_T}\right]=\mathbbm{E}\left[\left(\mathbf{k}^\prime-F_{_T}\mathbf{1}^\prime\right)^+\boldsymbol{\tilde\Omega}\left(F_{_T}\mathbf{1}-\mathbf{k}\right)^+\right]=0,
\end{equation*}
and therefore the fair-value swap rate becomes 
\begin{equation}\label{eq:zerointegrationerror}
\mathbbm{E}\left[\text{tr}\left(\boldsymbol\Omega^S\left(\mathbf{F}_{_T}-\mathbf{F}_{_0}\right)\left(\mathbf{F}_{_T}-\mathbf{F}_{_0}\right)^\prime\right)\right]=\mathbbm{E}\left[\text{tr}\left(\boldsymbol\Omega^S\mathbf{F}_{_T}\mathbf{F}_{_T}^\prime\right)\right]-\text{tr}\left(\boldsymbol\Omega^S\mathbf{F}_{_0}\mathbf{F}_{_0}^\prime\right)=-\mathbf{P}_{_0}^\prime\boldsymbol{\tilde\Omega}\mathbf{C}_{_0}.
\end{equation}
That is, the fixed leg can be derived from only the current prices $\mathbf{P}_{_0}$ and $\mathbf{C}_{_0}$ of traded vanilla options with strikes $\mathbf{k}$, without using the replication theorem of \cite{CM01}.

Now consider $d=1$ and $\boldsymbol{\tilde\Omega}=1$. Then $\mathbf{F}=\left(P,C\right)^\prime$ is the joint forward price process of a put and a call option with the same strike $k$, and the pay-off function becomes $\phi\left(\hat{\mathbf{z}}\right)=\hat{P}\hat{C}$. The fair-value swap rate is $\mathbbm{E}\left[\left(P_{_T}-P_{_0}\right)\left(C_{_T}-C_{_0}\right)\right]=-P_{_0}C_{_0}$. This swap can be hedged exactly by dynamically holding $P_{t-1}$ calls and $C_{t-1}$ puts from time $t-1$ to $t$, which corresponds to a straddle position.\footnote{To see this, consider the daily value increment of a straddle swap: $\mathbbm{E}_t\left[\sum_{\boldsymbol\Pi_{_D}}\hat{P}\hat{C}\right]-\mathbbm{E}_{t-1}\left[\sum_{\boldsymbol\Pi_{_D}}\hat{P}\hat{C}\right]
=\hat{P}_t\hat{C}_t+\mathbbm{E}_t\left[\left(P_{_T}-P_t\right)\left(C_{_T}-C_t\right)\right]-\mathbbm{E}_{t-1}\left[\left(P_{_T}-P_{t-1}\right)\left(C_{_T}-C_{t-1}\right)\right]
=-P_{t-1}\hat{C}_t-C_{t-1}\hat{P}_t
$, where all pay-offs prior to time $t-1$ cancel out and the argument from \eqref{eq:zerointegrationerror} applies to the expectations.}

\subsection{Frequency Swaps}

\gls{di} swap contracts allow buyers and sellers to hedge their exposure perfectly by trading in the underlying assets $\mathbf{F}$ whenever the swap is monitored. However, given transaction costs, it may be more practical for them to hedge at a lower frequency. Hedging may be based on some partition $\boldsymbol\Pi_h$ when the monitoring partition is $\boldsymbol\Pi_m\supset\boldsymbol\Pi_h$. For example, it may be convenient to buy a daily monitored swap and hedge once every month. In this case the residual exposure corresponds to a frequency swap with the floating leg
\begin{equation*}
\sum_{\boldsymbol\Pi_m}\phi\left(\mathbf{\hat{z}}\right)-\sum_{\boldsymbol\Pi_h}\phi\left(\mathbf{\hat{z}}\right).
\end{equation*}
The \gls{ap} implies $\mathbbm{E}\left[\sum_{\boldsymbol\Pi_m}\phi\left(\mathbf{\hat{z}}\right)\right]=\mathbbm{E}\left[\sum_{\boldsymbol\Pi_h}\phi\left(\mathbf{\hat{z}}\right)\right]=v^\phi_{_0}$ and, because the corresponding swap rates for the two floating components cancel out, the fair-value swap rate of this frequency swap is zero at inception. However, for $t>0$ the \gls{pnl} need not be zero in the presence of a hedging error. In fact, for $t\in\boldsymbol\Pi_h$ the mark-to-market \gls{pnl} on a \gls{di} frequency swap is
\begin{equation*}
\mathbbm{E}_t\left[\sum_{\boldsymbol\Pi_m}\phi\left(\mathbf{\hat{z}}\right)-\sum_{\boldsymbol\Pi_h}\phi\left(\mathbf{\hat{z}}\right)\right]=\sum_{\boldsymbol\Pi_m\cap[0,t]}\phi\left(\mathbf{\hat{z}}\right)-\sum_{\boldsymbol\Pi_h\cap[0,t]}\phi\left(\mathbf{\hat{z}}\right).
\end{equation*}
As long as the floating leg of a frequency swap depends only on the prices of traded contracts, e.g. for $\mathbf{z}=x$ and $\phi=\lambda$, pricing and hedging this frequency swap is exact.

\section{Empirical Study}\label{sec:empirical}\glsresetall

Here we analyse the historical performance of DI swap contracts on the \gls{snp} over an 18-year period from January 1996 to December 2013 using term-structure \gls{pnl} time series for different constant-maturities. These `unrealised' \glspl{pnl} are our empirical observations on the value increments of the price processes of the diverse swap contracts. 
In contrast to most previous studies, with the notable exception of \cite{KNS13}, we examine swaps with realised legs based on \gls{di} pay-offs. For the pricing of moment swaps, i.e. for determining their fair-value swap rates, we do not need to rely on market quotes which are not currently available in any case. Rather, we derive our fixed legs from vanilla \gls{otm} option prices and in the case of straddle swaps the fair values can be computed from the available traded strikes.

\subsection{Data and Methodology}\label{subsec:data}

Following \cite{CW09}, \cite{T10} and others we generate \glspl{pnl} as the difference between the observed floating pay-off under the physical measure and its synthetic fair value under the risk-neutral measure. 
%
We obtain daily closing prices $P_t$ and $C_t$ of all traded European put and call options on the S\&P 500 between January 1996 and December 2013 and follow the data filtering methodology, and the standardisation of moments described in \cite{RA16}.\footnote{The standardization  follows \cite{KNS13}.}  This way we eliminate unreliable prices,  preclude static arbitrage across strikes and maturity, and employ investable, constant maturity P\&L data.\footnote{Much other empirical work in on the swaps approach to variance risk premia, with the notable exception of \cite{ELW10}, fails in these properties. Either it constructs systematically-varying maturity data, derived from holding a swap until just before maturity the rolling to another swap with the same initial maturity, tracking observations on the realised pay-off and swap rate. Another alternative is to linearly interpolate synthetic constant-maturity swap rates and calculate the corresponding realised pay-off on every monitoring period. But this practice introduces  artefactual autocorrelation  when sampling P\&L at a higher frequency than the swap maturity. Also, \cite{CW09} and \cite{AB13}  examine risk premia that are not  investable.}


\subsection{ S\&P500 Risk Premia for DI Swaps}

\noindent The figures in this section depict the cumulative risk premia for constant-maturity moment swaps over the entire sample period. We examine their dependence on the maturity of the swap and  the monitoring frequency of the realised leg, which is the same as the rebalancing of the implied leg. 
In each case the total premia is disaggregated into realised and implied components, using Theorem 3. 
 
First we investigate the term-structure of higher-moment risk premia. Theorem 3 is applied to 30-, 90- and 180-day DI moment swap examples listed in Section \ref{sec:ms}, under daily monitoring. That is, we decompose the total P\&L into realised and implied components along the S\&P500 term structure. 
\begin{figure}[h!]
	\begin{centering}
		\includegraphics[width=1\linewidth,trim=3.05cm 0.25cm 2.15cm 0cm,clip=true]{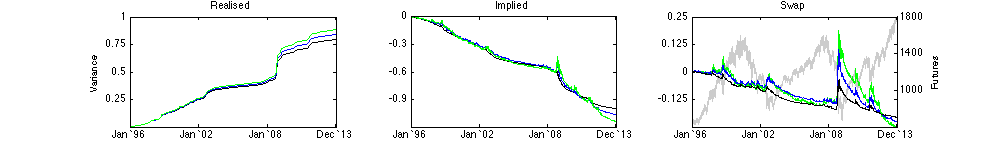}
		\includegraphics[width=1\linewidth,trim=3.05cm 0.25cm 2.15cm 0cm,clip=true]{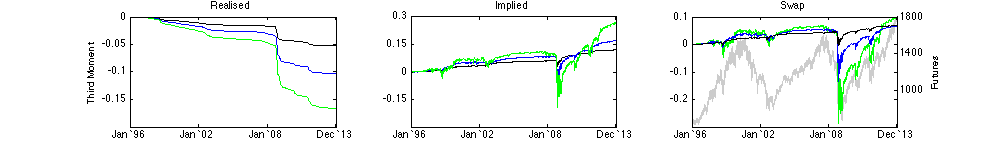}
		\includegraphics[width=1\linewidth,trim=3.05cm 0.25cm 2.15cm 0cm,clip=true]{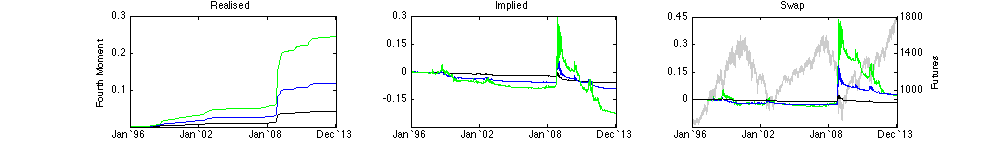}
		\includegraphics[width=1\linewidth,trim=3.05cm 0.25cm 2.15cm 0cm,clip=true]{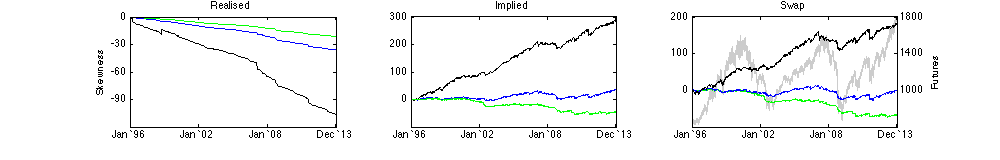}
		\includegraphics[width=1\linewidth,trim=3.05cm 0.25cm 2.15cm 0cm,clip=true]{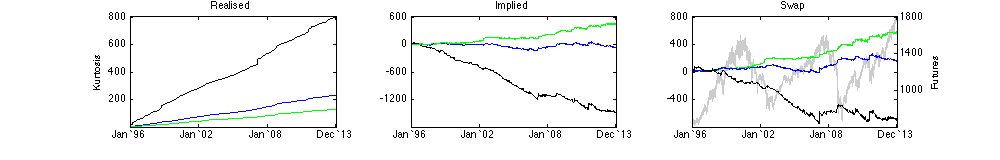}

		\par\end{centering}
	{\caption[Risk premium maturity time series]{\small Time series for daily-monitored 30-day (black), 90-day (blue) and 180-day (green)  cumulative moment risk premia. The secondary axis on the right refers to the 30-day S\&P500 forward contract plotted in grey. These graphs decompose the total cumulative risk premia into realised and implied components according to Equation \eqref{eq:realisedimplied}.}\label{fig:varianceswapgraphsmaturity}}
\end{figure}
Figure \ref{fig:varianceswapgraphsmaturity} depicts the results using a black line for the P\&L on  30-day DI moment swaps, blue for 90-day swaps and green for DI swaps with 180 days to maturity. Note that the realised components depend on maturity because the characteristics include contracts on options of that maturity. The skewness and kurtosis risk premia exhibit similar but opposite effects in both their implied and their realised components, both components become smaller in magnitude as maturity increases, and the implied component dominates the overall risk premium. The 30-day skew premium (black line) tends to be positive, except during turbulent market crises periods. The skew premium at 90 days (blue) is much smaller and close to zero and at 180 days  (green) it tends to be negative. Similar features are evident in the kurtosis premium but with opposite signs: it is typically negative at 30 days, but sharply increases during periods leading up to a market crisis. As expected, the kurtosis premium is near zero at longer maturity.
\begin{figure}[h!]
	\vspace{10pt}
	\begin{centering}
		\includegraphics[width=1\linewidth,trim=2.85cm 0.25cm 2.15cm 0cm,clip=true]{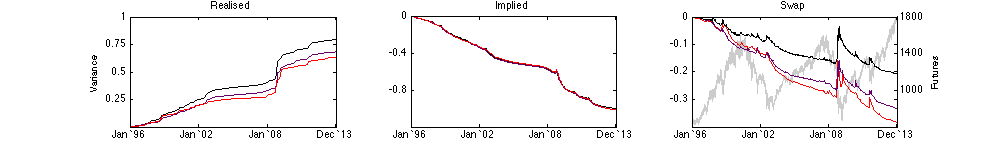}
		\includegraphics[width=1\linewidth,trim=2.85cm 0.25cm 2.15cm 0cm,clip=true]{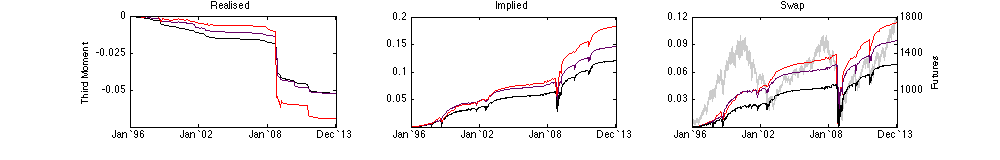}
		\includegraphics[width=1\linewidth,trim=2.85cm 0.25cm 2.15cm 0cm,clip=true]{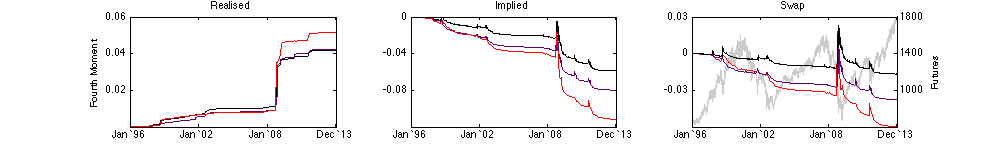}
			\includegraphics[width=1\linewidth,trim=2.85cm 0.25cm 2.15cm 0cm,clip=true]{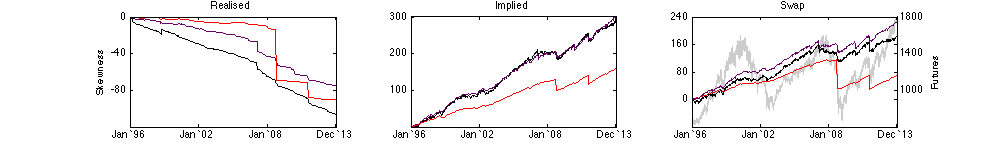}
			\includegraphics[width=1\linewidth,trim=2.85cm 0.25cm 2.15cm 0cm,clip=true]{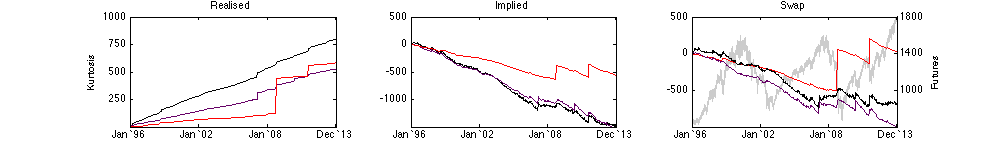}
		\par\end{centering}
	{ \caption[Risk premium frequency time series]{\small Time series of cumulative 30-day  variance,  third-moment and fourth-moment risk premia 
			based on daily (black), weekly (purple) and monthly (red) monitoring. The secondary axis on the right refers to the 30-day S\&P500 forward contract plotted in grey. These graphs decompose the total cumulative risk premia into realised and implied components according to Equation \eqref{eq:realisedimplied}.}\label{fig:varianceswapgraphsfrequency}}
\end{figure}

Figure \ref{fig:varianceswapgraphsfrequency} presents cumulative 30-day higher-moment risk premia when the realised characteristic is monitored at different frequencies. The implied component of the variance risk premium does not depend on the monitoring frequency.\footnote{That is, when the replication basket of options is rebalanced daily to constant 30-day maturity and valued by marking-to-market (i.e. the black line), the cumulative change in the implied component is approximately the same as if the rebalancing and valuing happens weekly (purple) or monthly (red).} The very small variation evident in the top centre graph is due to variation in the separation strike of the replication portfolio. It is the realised leg which drives the dependence of the variance premium on the monitoring frequency. Overall, it becomes smaller and less variable as monitoring frequency increases.\footnote{
	Theoretical results to support these observations are model dependent. For instance, when $dS_t=\mu S_t + \sigma S_t dW_t$ where $W_t$ is a Brownian motion it is straightforward to show that the risk premium associated with the conventional realised variance over a regular partition of $[0,T]$ into $N$ elements is $\mu\left(\mu-\sigma^2\right)T^2N^{-1}$ and the variance of this realised variance is $2\sigma^4T^2N^{-1}+ 4\mu^2\sigma^2T^3N^{-2}$. Further model-dependent results, available from the authors on request,  confirm the statement for some other processes and DI variance characteristics.} It is usually negative but during  the collapse of Lehman Brothers in September 2008 and in August 2011 at the onset of the European sovereign debt crisis it is, briefly, highly positive. 

By contrast, the third-moment premium is usually positive, but falls sharply during crisis periods when the negative skew in realised returns on equities becomes especially pronounced. This is driven by the large jump down in the realised component during September 2008 (left-hand graph in the second row). More generally this premium is dominated by the implied component  depicted in the centre graph. The effect of rebalancing the separation strike is more evident here than it is in the implied variance. For instance, in the monthly-monitored (red) time series the failure to rebalance the separation strike every day implies using higher-priced in-the-money calls in the replication portfolio during an upwards trending market, or higher-priced in-the-money puts in the replication portfolio during a downward market. A similar but opposite effect is evident in the implied component of the fourth-moment risk premium. As expected, given that the fourth moment captures outliers in a distribution, this    premium is dominated by jumps in the index and is strongly positive during  crisis periods.

\subsection{Risk Premia on Calendar, Frequency and Straddle Swaps}
\noindent Given that risk premia can exhibit a strong term-structure pattern, as in Figure \ref{fig:varianceswapgraphsmaturity}, systematic risk premia could be traded by entering a floating-floating `calendar swap' which exchanges two realised characteristics, monitored at the same frequency, but with different maturities. For instance, a 180/30-day calendar variance swap would pay the forward realised variance, from 30 days after inception of the contract up to 180 days, in exchange for the corresponding fair-value swap rate, which equals the difference between the 180-day and 30-day swap rates. 

\begin{table}[h!]\small
\vspace{10pt}
\begin{center}
\begin{tabular}{ll|ccccc:ccc}
Calendar & & $V^{(2)}$ & $V^{(3)}$ & $V^{(\bar{3})}$ & $V^{(4)}$ & $V^{(\bar{4})}$ & $V^{[k_1]}$ & $V^{[k_2]}$ & $V^{[k_3]}$ \\\hline
\multirow{3}{*}{\vtop{\hbox{\strut $[\tau=180]$}\hbox{\strut --$[\tau=30]$}}} & $\boldsymbol\Pi_{_D}$ & -0.05 & 0.02 & -1.30 & 0.01 & 1.12 & 0.16 & 0.18 & 0.20 \\
& $\boldsymbol\Pi_{_W}$ & -0.03 & 0.02 & -1.54 & 0.04 & 1.20 & 0.25 & 0.22 & 0.20 \\
& $\boldsymbol\Pi_{_M}$ & -0.02 & 0.10 & -0.18 & -0.08 & -0.02 & 0.05 & 0.18 & 0.12\\
\multicolumn{10}{c}{}\\
Frequency & & $V^{(2)}$ & $V^{(3)}$ & $V^{(\bar{3})}$ & $V^{(4)}$ & $V^{(\bar{4})}$ & $V^{[k_1]}$ & $V^{[k_2]}$ & $V^{[k_3]}$ \\\hline
$\tau=30$ & & -0.63 & 0.37 & -0.66 & -0.41 & 0.59 & 0.27 & 0.16 & 0.45 \\
$\tau=90$ & $\boldsymbol\Pi_{_M}-\boldsymbol\Pi_{_D}$ & -0.52 & 0.53 & 0.31 & -0.54 & -0.11 & 0.37 & 0.30 & 0.28 \\
$\tau=180$ & & -0.46 & 0.48 & 1.60 & -0.61 & -1.77 & -0.09 & -0.04 & 0.07 
\end{tabular}
\end{center}
\vspace{-10pt}
{\caption[Risk premia]{\small Standardised risk premia between January 1996 and December 2013 on daily, weekly and monthly monitored 180-for-30-day calendar swaps (above) and 30-day, 90-day and 180-day constant-maturity monthly-daily frequency swaps (below), where the swap rates are exchanged for: moment swaps on the log price $V^{(n)}$, the skewness swap $V^{(\bar{3})}$, the kurtosis swap $V^{(\bar{4})}$ as well as straddle swaps with strikes $k_1=1000$, $k_2=1100$ and $k_3=1200$.}\label{tab:frequencycalendarswapmeans}}
\end{table}

Table \ref{tab:frequencycalendarswapmeans} summarises the risk premia on some floating-floating swaps. For ease of comparison each premium is standardized by dividing by its standard deviation and annualising. The top panel exhibits the standardised risk premia obtained on 180-for-30-day calendar swaps monitored at three different frequencies. 
 As expected from the very different features of the skewness and kurtosis risk premia displayed in Figure \ref{fig:varianceswapgraphsmaturity}, the  skewness (kurtosis) calendar swaps  exhibit large negative (positive)  premia at the daily and weekly monitoring frequencies. No other calendar swaps display significant results.

The lower panel in Table \ref{tab:frequencycalendarswapmeans} reports the standardized risk premia on `frequency swaps' which exchange two realised legs of the same maturity that are monitored at different frequencies. For instance, a monthly-daily variance frequency swap receives monthly and pays daily realised variance. Conveniently, the AP implies that the fair-value rate on this type of swap is zero, by definition, but the risk premium may be positive or negative depending on the sample period and underlying characteristic. 
These frequency swaps tend to give larger risk premia in general and the skewness and kurtosis frequency swaps in particular have large risk premia ($1.60$ and $-1.77$ respectively) at the 180-day maturity.

\begin{figure}[h!]
\begin{centering}
\includegraphics[width=1\linewidth,trim=3.35cm 0.75cm 2.15cm 0cm,clip=true]{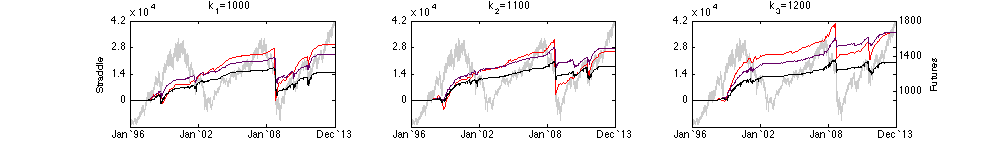}
\par\end{centering}
{ \caption[Straddle swaps time series]{\small Time series for the cumulative risk premia on  30-day constant-maturity straddle swaps with strikes $k_1=1000$, $k_2=1100$ and $k_3=1200$, denoted by $V^{[k_1]}$, $V^{[k_2]}$ and $V^{[k_3]}$ and defined as in the previous section. Black, purple and red lines refer to swaps with realised characteristics that are monitored on a daily, weekly and monthly basis, respectively. Since the implied leg of a straddle swap is always zero, the risk premium is driven entirely by the realised component.}
\label{fig:straddleswapgraphs}}
\end{figure}

Figure \ref{fig:straddleswapgraphs} depicts the time series of risk premia on straddle swaps with strikes $k_1=1000$, $k_2=1100$ and $k_3=1200$ when monitored at different frequencies.\footnote{The choice of strike here allows us to investigate the behaviour of the swaps over the 18-year sample period because call and put options at these strikes were traded most of the time. We exclude strangle swaps from this analysis since they are more expensive to trade, due to the concentration of liquidity at the money, but results are available from the authors on request.} The risk premium on these swaps can be large and negative during a crisis, e.g. in September 2008 and August 2011. Otherwise, the risk premium is small and positive, and it is greater for straddle swaps that are monitored weekly or monthly than for those that are monitored daily.

\section{Conclusions}\label{sec:conc}\glsresetall

Fair-value rates for conventional variance swaps are biased due to discrete-monitoring, jump and numerical integration errors. As a result market rates can deviate substantially from their fair values, especially during turbulent periods. This has been a catalyst for much recent research on finding arbitrage bounds for these errors. Another, very original strand of research, pioneered by \cite{N12} and developed by \cite{B14}, suggests different definitions for the realised variance for which more precise fair values may be obtained. Our research develops this second strand to derive a general theory for variance, higher-moment and other so-called \gls{di} pay-offs for which exact fair values are derived in a totally model-free setting.

By restricting the aggregation property to characteristics and processes which unify the two definitions of  \cite{N12} and \cite{B14} we have followed the lead in the concluding remarks in \cite{N12} to characterise a vector space of what we now term 'discretisation-invariant' \gls{di} pay-offs. Theorem 1 characterises all twice-continuously differentiable pay-off functions $\phi$ having this property as solutions to a second order system of partial differential equations. Theorem 2 focusses on a particular sub-class, i.e. those for which the pay-off is analytic.  Theorem 3 shows how the value of these swaps can be replicated by dynamically rebalancing portfolios of the underlying and certain fundamental contracts, and then we consider some special \gls{di} pay-offs which correspond to higher-order moments of a univariate distribution, and floating-floating swaps associated with different monitoring partitions, which have zero fair value. These DI swaps also identify the residual hedging risk when the replication portfolio is rebalanced at a frequency other than the monitoring one. 

\gls{di} variance swaps have several advantages over conventional variance swaps: (i) there is no jump or other model dependence error in their theoretical fair-value swap rate; consequently (ii) issuers would face smaller residual hedging risks; and (iii) the absence of arbitrage should yield market prices that are within the bid-ask spread of the fair-value, since the only approximation errors arise from numerical integration; and (iv) issuers would have greater flexibility to choose the monitoring frequency of the realised leg because the fair-value swap rate is the same for all frequencies, the monitoring does not even need to be regular. All these advantages also apply to higher-order moment risk premia.   

The calculation of the fair-value for a DI moment swap is still subject to a computation error because their replication requires numerical integration over option prices at traded strikes to approximate an integral formula. However, a sub-space of DI pay-offs can be defined for which even this  error is zero. These swaps have pay-off functions defined by bi-linear forms of traded call and put prices. Again, an infinite variety of such SDI pay-offs exists and we have only investigated so-called `straddle swaps' empirically. Their fair-value rates are derived from the product of current put and call prices with the same strike.

We believe that the concepts and empirical work presented in this paper will lay the foundations for  research into new  sources of risk which become tradable via \gls{di} pay-offs. Further empirical studies might consider multivariate underlying for these swaps (e.g. swaps on realised joint pay-offs of S\&P 500 and VIX futures, and the addition of foreign exchange rates). This could open new strands for research on correlation and covariance swaps, and on currency-protected products. More generally, we could investigate moments of univariate and multivariate distributions based on other equity indices, or  bond and commodity index futures. Further empirical work would also be interesting on other DI pay-offs not linked to moments, especially those without numerical integration error, and frequency and calendar swaps which trade on the term structures of the realised and implied legs, respectively.

Further empirical work on swaps that are monitored at irregular frequencies might include deriving a variance risk premium from a realised pay-off function that is monitored in transaction time. Such a swap could be monitored whenever cumulative trading in the underlying reaches a pre-defined level. The S\&P 500 `transaction time' variance risk premium will be much less volatile than the standard variance swap rate, so banks would take much less risk by paying these rather than swaps based on the standard realised variance.  Finally, it would be interesting for hedge funds and other investors with relatively short-term horizons to construct optimal portfolios which diversify  variance risk through higher-moment DI swaps. 

\newpage
\singlespacing
\bibliographystyle{plainnat}
\bibliography{bibliographyV8}

\doublespacing
\begin{appendix}

\setlength\abovedisplayskip{6pt}
\setlength\belowdisplayskip{6pt}
\setlength\abovedisplayshortskip{0pt}
\setlength\belowdisplayshortskip{6pt}

\newpage

\section{Theoretical Appendix}\label{sec:app1}\glsresetall
Let $\mathbf{F}$ be a multivariate $\mathbbm{Q}$-martingale and set $\mathbf{x}:= \ln \mathbf{F}$. Let  $\varphi^\star:\mathbbm{R}^n\times\mathbbm{R}^n\rightarrow\mathbbm{R}$ denote a pay-off function on $\left(\mathbf{x},\mathbf{x}+\mathbf{\hat{x}}\right)$. The aggregation property as introduced by \cite{B14} may then be written:
\begin{equation}\label{eq:aggregationpropertybondarenko}
\mathbbm{E}\left[\sum_{\boldsymbol\Pi_{_N}}\varphi^\star\left(\mathbf{x},\mathbf{x}+\mathbf{\hat{x}}\right)\right]=\mathbbm{E}\left[\varphi^\star\left(\mathbf{x}_{_0},\mathbf{x}_{_T}\right)\right]\quad \forall \,\,\mbox{partitions}\,\, \boldsymbol\Pi_{_N}.
\end{equation}

\noindent {\bf Lemma 1:}  When \eqref{eq:aggregationproperty} is applied to the adapted process $\mathbf{z}=\left(\mathbf{F},\mathbf{x}\right)^\prime$ with $\mathbf{x}:=\ln\mathbf{F}$, the properties \eqref{eq:aggregationpropertybondarenko} and \eqref{eq:aggregationproperty} are equivalent.\vspace{12pt}

\noindent {\bf Proof:} Note that $\mathbf{\hat{x}}=\ln\left(\mathbf{F}+\mathbf{\hat{F}}\right)-\ln\mathbf{F}$ and 
$\mathbf{F}=\left(\mathrm{e}^{\mathbf{\hat{x}}}-1\right)^{-1}\mathbf{\hat{F}}$, where all vector operations are understood component-wise. Then $\varphi\left(\mathbf{\hat{z}}\right)=\varphi^\star\left(\mathbf{x},\mathbf{x}+\mathbf{\hat{x}}\right)$ and $\varphi\left(\mathbf{z}_{_T}-\mathbf{z}_{_0}\right)=\varphi^\star\left(\mathbf{x}_{_0},\mathbf{x}_{_T}\right)$ in particular.\qed\vspace{12pt}

\noindent {\bf Lemma 2:} The \gls{ap} is necessary for the discrete monitoring error \eqref{eq:fdiscretemonitoringerror} to equal zero, i.e.
\begin{equation}
\mathbbm{E}\left[\sum_{\boldsymbol\Pi_{_N}}\phi\left(\mathbf{\hat{z}}\right)\right]=\mathbbm{E}\left[\langle\mathbf{z}\rangle^\phi_{_T}\right]\quad \forall \,\, \boldsymbol\Pi_{_N}.
\end{equation}
Furthermore, if $\lim_{\boldsymbol\Pi_{_N}\rightarrow\boldsymbol\Pi}\mathbbm{E}\left[\sum_{\boldsymbol\Pi_{_N}}\phi\left(\mathbf{\hat{z}}\right)\right]=\mathbbm{E}\left[\langle\mathbf{z}\rangle^\phi_{_T}\right]$ the \gls{ap} is also sufficient. 
\vspace{12pt}

\noindent {\bf Proof:} If \eqref{eq:nodiscretemonitoringerror} holds for any partition it must hold for $\boldsymbol\Pi_{_N}$ as well as for the trivial partition $[0,T]$ in particular. Then $\mathbbm{E}\left[\sum_{\boldsymbol\Pi_{_N}}\phi\left(\mathbf{\hat{z}}\right)\right]=\mathbbm{E}\left[\phi\left(\mathbf{z}_{_T}-\mathbf{z}_{_0}\right)\right]$. 
Taking the limit as $\boldsymbol\Pi_{_N}\rightarrow\boldsymbol\Pi$ yields the equivalence.\qed\\

\subsection{Proof of Theorem 1}\label{app:theorem1}

Let the forward price process $\mathbf{F}$ follow the $\mathbbm{Q}$-dynamics $d\mathbf{F}_t=\boldsymbol\sigma_td\mathbf{W}_t$ where $\mathbf{\boldsymbol\sigma}=\left\{\boldsymbol\sigma_t\right\}_{t\in\boldsymbol\Pi}\in\mathbbm{R}^{d\times d}$ and $\mathbf{W}=\left\{\mathbf{W}_t\right\}_{t\in\boldsymbol\Pi}\in\mathbbm{R}^d$ is a multivariate Wiener process with $T^{-1}\langle\mathbf{W}\rangle_t=\mathbf{I}$, the identity matrix. Then $d\langle\mathbf{F}\rangle_t=\boldsymbol\sigma_t\boldsymbol\sigma_t^\prime dt$ is the quadratic covariation process of $\mathbf{F}$.\footnote{The quadratic covariation is a straightforward generalisation of the quadratic variation for multivariate processes and is defined as $\langle\mathbf{z}\rangle_{_T}:=\lim_{\boldsymbol\Pi_{_N}\rightarrow\boldsymbol\Pi}\sum_{\boldsymbol\Pi_{_N}}\mathbf{\hat{z}}_i\mathbf{\hat{z}}_i^\prime=\int_{\boldsymbol\Pi}d\mathbf{z}_td\mathbf{z}_t^\prime$. Note that the quadratic covariation $\langle\mathbf{z}\rangle$ is a matrix while the $\phi$-variation $\langle\mathbf{z}\rangle^\phi$ is a scalar.} 
Let $\boldsymbol\Delta:=\nabla^\prime_{_\mathbf{F}}\mathbf{z}\in\mathbbm{R}^{n\times d}$ and $\boldsymbol\Gamma:=\nabla^{\prime\prime}_{_\mathbf{F}}\boldsymbol\Delta\in\mathbbm{R}^{n\times d\times d}$ denote the first and second partial derivatives of $\mathbf{z}$ w.r.t. $\mathbf{F}$ where $\nabla_{_\mathbf{F}}:=\left(\tfrac{\partial}{\partial F_1},\ldots,\tfrac{\partial}{\partial F_d}\right)^\prime$.
Then, applying It\^o's Lemma and the cyclic property of the trace operator, we have
\begin{equation}\label{eq:derivativesdynamics}
d\mathbf{z}_t=\boldsymbol\Delta_td\mathbf{F}_t+\tfrac{1}{2}\text{tr}\left(\boldsymbol\Gamma_td\left\langle\mathbf{F}\right\rangle_t\right),
\end{equation}
so that the quadratic covariation process of $\mathbf{z}$ follows the dynamics
\begin{equation}\label{eq:quadcovar}
d\langle\mathbf{z}\rangle_t=\boldsymbol\Delta_t\boldsymbol\sigma_t\boldsymbol\sigma_t^\prime\boldsymbol\Delta^\prime_tdt.
\end{equation}
Since we want the discrete monitoring error to be zero for all possible forward price processes, it must hold in particular for any specific martingale. We can therefore derive a necessary condition for the functions spanning $\mathbbm{F}$ by starting from the assumptions that \eqref{eq:nodiscretemonitoringerror} holds w.r.t. $\left(\phi,\mathbf{z}\right)$ and that $\mathbf{z}$ follows the dynamics specified in \eqref{eq:derivativesdynamics}.

Denote the Jacobian vector of first partial derivatives of $\phi$ by $\mathbf{J}\left(\mathbf{\hat{z}}\right):=\nabla_{_\mathbf{z}}\phi\left(\mathbf{\hat{z}}\right)\in\mathbbm{R}^{n}$ and the Hessian matrix of second partial derivatives of $\phi$ by $\mathbf{H}\left(\mathbf{\hat{z}}\right):=\nabla^\prime_{_\mathbf{z}}\mathbf{J}\left(\mathbf{\hat{z}}\right)\in\mathbbm{R}^{n\times n}$ where $\nabla_{_\mathbf{z}}:=\left(\tfrac{\partial}{\partial\hat{z}_1},\ldots,\tfrac{\partial}{\partial\hat{z}_n}\right)^\prime$.
Then It\^o's Lemma yields
\begin{equation}\label{eq:itototalincrement}
\phi\left(\mathbf{z}_{_T}-\mathbf{z}_{_0}\right)=\int_{\boldsymbol\Pi}\mathbf{J}^\prime\left(\mathbf{z}_t-\mathbf{z}_{_0}\right)d\mathbf{z}_t+\tfrac{1}{2}\text{tr}\int_{\boldsymbol\Pi}\mathbf{H}\left(\mathbf{z}_t-\mathbf{z}_{_0}\right)d\langle\mathbf{z}\rangle_t.
\end{equation}
Similarly,
\begin{eqnarray}\label{eq:itopartialincrement}
\sum_{\boldsymbol\Pi_{_N}}\phi\left(\mathbf{\hat{z}}_i\right)&=&\sum_{i=1}^N\left\{\int_{t_{i-1}}^{t_i}\mathbf{J}^\prime\left(\mathbf{z}_t-\mathbf{z}_{t_{i-1}}\right)d\mathbf{z}_t+\tfrac{1}{2}\text{tr}\int_{t_{i-1}}^{t_i}\mathbf{H}\left(\mathbf{z}_t-\mathbf{z}_{t_{i-1}}\right)d\langle\mathbf{z}\rangle_t\right\}\nonumber\\
&=&\int_{\boldsymbol\Pi}\mathbf{J}^\prime\left(\mathbf{z}_t-\mathbf{z}_{m(t)}\right)d\mathbf{z}_t+\tfrac{1}{2}\text{tr}\int_{\boldsymbol\Pi}\mathbf{H}\left(\mathbf{z}_t-\mathbf{z}_{m(t)}\right)d\langle\mathbf{z}\rangle_t,
\end{eqnarray}
where $m(t):=\max\{t_i\in\boldsymbol\Pi_{_N}|t_i\le t\}$. Taking the limit as $\boldsymbol\Pi_{_N}\rightarrow\boldsymbol\Pi$ yields the $\phi$-variation
\begin{equation}\label{eq:itofvariation}
\langle\mathbf{z}\rangle_{_T}^\phi=\int_{\boldsymbol\Pi}\mathbf{J}^\prime d\mathbf{z}_t+\tfrac{1}{2}\text{tr}\int_{\boldsymbol\Pi}\mathbf{H}d\langle\mathbf{z}\rangle_t,
\end{equation}
where $\mathbf{J}:=\mathbf{J}\left(\mathbf{0}\right)$ and $\mathbf{H}:=\mathbf{H}\left(\mathbf{0}\right)$. With \eqref{eq:itototalincrement} and \eqref{eq:itofvariation}, the condition \eqref{eq:nodiscretemonitoringerror} is equivalent to 
\begin{equation}\label{eq:expectation}
\mathbbm{E}\left[\int_{\boldsymbol\Pi}\left[\mathbf{J}\left(\mathbf{z}_t-\mathbf{z}_{_0}\right)-\mathbf{J}\right]^\prime d\mathbf{z}_t+\tfrac{1}{2}\text{tr}\int_{\boldsymbol\Pi}\left[\mathbf{H}\left(\mathbf{z}_t-\mathbf{z}_{_0}\right)-\mathbf{H}\right]d\langle\mathbf{z}\rangle_t\right]=0.
\end{equation}
Substituting \eqref{eq:derivativesdynamics} and \eqref{eq:quadcovar} in \eqref{eq:expectation}, and using $\mathbbm{E}\left[d\mathbf{F}_t\right]=0$ yields that \eqref{eq:nodiscretemonitoringerror} is equivalent to 
\begin{equation}\label{eq:itocondition}
\text{tr}\mathbbm{E}\left[\int_{\boldsymbol\Pi}\left\{\left[\mathbf{J}\left(\mathbf{z}_t-\mathbf{z}_{_0}\right)-\mathbf{J}\right]^\prime\boldsymbol\Gamma_t+\boldsymbol\Delta^\prime_t\left[\mathbf{H}\left(\mathbf{z}_t-\mathbf{z}_{_0}\right)-\mathbf{H}\right]\boldsymbol\Delta_t\right\}\boldsymbol\sigma_t\boldsymbol\sigma_t^\prime dt\right]=0.
\end{equation}
Now consider the spectral decomposition 
\begin{equation}\label{eq:spectraldecomposition}
\left[\mathbf{J}\left(\mathbf{z}_t-\mathbf{z}_{_0}\right)-\mathbf{J}\right]^\prime\boldsymbol\Gamma_t+\boldsymbol\Delta^\prime_t\left[\mathbf{H}\left(\mathbf{z}_t-\mathbf{z}_{_0}\right)-\mathbf{H}\right]\boldsymbol\Delta_t=:\mathbf{E}_t\mathbf{\Lambda}_t\mathbf{E}_t^\prime,
\end{equation}
where $\mathbf{\Lambda}_t=\text{diag}\left\{\lambda_{1t},\ldots,\lambda_{dt}\right\}$ is a diagonal matrix of eigenvalues and $\mathbf{E}_t$ is an orthogonal matrix of eigenvectors. In order to derive a necessary condition for \eqref{eq:nodiscretemonitoringerror} we select the particular volatility process:
\begin{equation*}
\boldsymbol\sigma_t:=\exp\left\{\tfrac{1}{2}\xi\mathbf{E}_t\mathbf{\Lambda}_t\mathbf{E}^\prime_t\right\},
\end{equation*}
where $\xi\in\mathbbm{R}$ is an arbitrary constant. Because $\exp\left\{\mathbf{E}\mathbf{\Lambda}\mathbf{E}^{-1}\right\}=\mathbf{E}\exp\left\{\mathbf{\Lambda}\right\}\mathbf{E}^{-1}$ for $\mathbf{\Lambda},\mathbf{E}\in\mathbbm{R}^{d\times d}$ 
we have
\begin{equation}\label{eq:processvariation}
\boldsymbol\sigma_t\boldsymbol\sigma_t^\prime=\mathbf{E}_t\exp\left\{\xi\mathbf{\Lambda}_t\right\}\mathbf{E}^\prime_t.
\end{equation}
Inserting \eqref{eq:spectraldecomposition} and again \eqref{eq:processvariation} into \eqref{eq:itocondition} and differentiating w.r.t. $T$, then using the cyclic property of the trace yields
\begin{equation*}
\mathbbm{E}\left[\text{tr}\left(\mathbf{\Lambda}_t\exp\left\{\xi\mathbf{\Lambda}_t\right\}\right)\right]=0.
\end{equation*}
Differentiating once w.r.t. $\xi$ and evaluating the equation at $\xi=0$ yields the condition
\begin{equation*}\label{eq:momentgeneratingfunction}
\mathbbm{E}\left[\text{tr}\left(\mathbf{\Lambda}_t^2\right)\right]=\sum_{i=1}^d\mathbbm{E}\left[\left(\lambda^i_t\right)^2\right]=0,
\end{equation*}
which implies that all eigenvalues in $\boldsymbol\Lambda_t$ must be equal to zero. Hence we know that both sides in \eqref{eq:spectraldecomposition} are zero and, given that this must hold for all $\mathbf{F}_t$ and $\mathbf{z}_{_0}$, we have
\begin{equation}\label{eq:maincondition}
\left[\mathbf{J}\left(\mathbf{\hat{z}}\right)-\mathbf{J}\right]^\prime\boldsymbol\Gamma+\boldsymbol\Delta^\prime\left[\mathbf{H}\left(\mathbf{\hat{z}}\right)-\mathbf{H}\right]\boldsymbol\Delta=\mathbf{0},
\end{equation}
where $\mathbf{F}$ and $\mathbf{\hat{z}}$ are independent variables. We have derived this $d\times d$ system of partial differential equations based on the assumption that $\mathbf{F}$ follows a particular martingale diffusion, so it represents a necessary condition for the more general case where $\mathbf{F}$ can be any martingale diffusion. The two conditions are equivalent since \eqref{eq:maincondition} is also sufficient for \eqref{eq:itocondition} to hold.\footnote{The proof  can be performed analogously, this time assuming the \gls{ap}, by substituting \eqref{eq:itototalincrement} and \eqref{eq:itopartialincrement} into condition \eqref{eq:aggregationproperty} which yields the same solution \eqref{eq:maincondition}. This version does not require the existence of the $\phi$-variation. Furthermore, if we relax our assumption that $\mathbf{F}$ follows a diffusion and allow any martingale then \eqref{eq:maincondition} still represents a necessary condition for \eqref{eq:itocondition}.}\qed

\subsection{Proof of Theorem 2}\label{app:theorem2}

When $\mathbf{z}=\left(\mathbf{F},\mathbf{x}\right)^\prime$ we have
$\boldsymbol\Delta(\mathbf{F})=\left(\mathbf{I},\text{diag}(\mathbf{F})^{-1}\right)^\prime\in\mathbbm{R}^{2d\times d}$ and $\boldsymbol\Gamma(\mathbf{F})=\left(\mathbf{0},-\text{diag}_3(\mathbf{F})^{-2}\right)^\prime\in\mathbbm{R}^{2d\times d\times d}$ where $\text{diag}_3(\mathbf{F})$ denotes a three dimensional tensor with the elements of $\mathbf{F}$ on the diagonal and zeros everywhere else. We shall further use the following decompositions:
\begin{equation*}
\left[\mathbf{J}\left(\mathbf{\hat{z}}\right)-\mathbf{J}\left(\mathbf{0}\right)\right]=\left(\begin{array}{c}\mathbf{J}_{_\mathbf{F}}\left(\mathbf{\hat{z}}\right)\\\mathbf{J}_{_\mathbf{x}}\left(\mathbf{\hat{z}}\right)\end{array}\right)\in\mathbbm{R}^{2d},
\end{equation*}
and
\begin{equation*}
\left[\mathbf{H}\left(\mathbf{\hat{z}}\right)-\mathbf{H}\left(\mathbf{0}\right)\right]=\left(\begin{array}{cc}\mathbf{H}_{_\mathbf{F}}\left(\mathbf{\hat{z}}\right)&\mathbf{G}\left(\mathbf{\hat{z}}\right)\\\mathbf{G}\left(\mathbf{\hat{z}}\right)^\prime&\mathbf{H}_{_\mathbf{x}}\left(\mathbf{\hat{z}}\right)\end{array}\right)\in\mathbbm{R}^{2d\times 2d}.
\end{equation*}
Then \eqref{eq:maincondition} may be written:
\begin{eqnarray*}
&&-\mathbf{J}_{_\mathbf{x}}\left(\mathbf{\hat{z}}\right)^\prime\text{diag}_3(\mathbf{F})^{-2}+\mathbf{H}_{_\mathbf{F}}\left(\mathbf{\hat{z}}\right)+\mathbf{G}\left(\mathbf{\hat{z}}\right)\text{diag}(\mathbf{F})^{-1}\\
&&+\text{diag}(\mathbf{F})^{-1}\mathbf{G}\left(\mathbf{\hat{z}}\right)^\prime+\text{diag}(\mathbf{F})^{-1}\mathbf{H}_{_\mathbf{x}}\left(\mathbf{\hat{z}}\right)\text{diag}(\mathbf{F})^{-1}=\mathbf{0},
\end{eqnarray*}
and multiplying from left and right with $\text{diag}(\mathbf{F})$ (note that $\mathbf{F}>\mathbf{0}$) yields
\begin{eqnarray*}
&&-\text{diag}(\mathbf{J}_{_\mathbf{x}}\left(\mathbf{\hat{z}}\right))+\text{diag}(\mathbf{F})\mathbf{H}_{_\mathbf{F}}\left(\mathbf{\hat{z}}\right)\text{diag}(\mathbf{F})\\
&&+\text{diag}(\mathbf{F})\mathbf{G}\left(\mathbf{\hat{z}}\right)+\mathbf{G}\left(\mathbf{\hat{z}}\right)^\prime\text{diag}(\mathbf{F})+\mathbf{H}_{_\mathbf{x}}\left(\mathbf{\hat{z}}\right)=\mathbf{0}.
\end{eqnarray*}
Since this condition must be fulfilled for all martingale It\^o processes $\mathbf{F}$ (and for $\mathbf{F}=\mathbf{1}$ in particular) this implies $\mathbf{H}_{_\mathbf{F}}\left(\mathbf{\hat{z}}\right)=\mathbf{0}$ and $\mathbf{G}\left(\mathbf{\hat{z}}\right)=\mathbf{0}$ (the latter because of symmetry of the Hessian matrix) as well as $\mathbf{H}_{_\mathbf{x}}\left(\mathbf{\hat{z}}\right)=\text{diag}(\mathbf{J}_{_\mathbf{x}}\left(\mathbf{\hat{z}}\right))$. Therefore the solution must take the form
\begin{equation*}
\phi\left(\mathbf{\hat{z}}\right)=\boldsymbol\alpha^\prime\mathbf{\hat{F}}+\text{tr}\left(\boldsymbol\Omega\mathbf{\hat{F}}\mathbf{\hat{F}}^\prime\right)+\boldsymbol\beta^\prime\left(\mathrm e^{\mathbf{\hat{x}}}-\mathbf{1}\right)+\boldsymbol\gamma^\prime\mathbf{\hat{x}},
\end{equation*}
where $\boldsymbol\alpha,\boldsymbol\beta,\boldsymbol\gamma\in\mathbbm{R}^{d}$ and $\boldsymbol\Omega^\prime=\boldsymbol\Omega\in\mathbbm{R}^{d\times d}$ is a symmetric matrix.

Swaps associated with $\boldsymbol\alpha$ are DI since $\lim_{\boldsymbol\Pi_N\rightarrow\boldsymbol\Pi}\sum_{\boldsymbol\Pi_{_N}}\mathbf{\boldsymbol\alpha}^\prime\mathbf{\hat{F}}_i=\mathbf{\boldsymbol\alpha}^\prime\left(\mathbf{F}_{_T}-\mathbf{F}_{_0}\right)$ even without expectation for any process. The same holds for swaps associated with $\boldsymbol\gamma$. For the swaps associated with $\boldsymbol\Omega$ we can apply
\begin{eqnarray*}
\mathbbm{E}\left[\lim_{\boldsymbol\Pi_N\rightarrow\boldsymbol\Pi}\sum_{\boldsymbol\Pi_{_N}}\text{tr}\left(\boldsymbol\Omega\mathbf{\hat{F}}_i\mathbf{\hat{F}}_i^\prime\right)\right]
&=&\mathbbm{E}\left[\text{tr}\left(\boldsymbol\Omega\lim_{\boldsymbol\Pi_N\rightarrow\boldsymbol\Pi}\sum_{\boldsymbol\Pi_{_N}}\left(\mathbf{F}_{t_i}-\mathbf{F}_{t_{i-1}}\right)\left(\mathbf{F}_{t_i}-\mathbf{F}_{t_{i-1}}\right)^\prime\right)\right]\\
&=&\mathbbm{E}\left[\text{tr}\left(\boldsymbol\Omega\lim_{\boldsymbol\Pi_N\rightarrow\boldsymbol\Pi}\sum_{\boldsymbol\Pi_{_N}}\left[\mathbf{F}_{t_i}\mathbf{F}_{t_i}^\prime-\mathbf{F}_{t_{i-1}}\mathbf{F}_{t_{i-1}}^\prime\right]\right)\right]\\
&=&\mathbbm{E}\left[\text{tr}\left(\boldsymbol\Omega\left[\mathbf{F}_{_T}\mathbf{F}_{_T}^\prime-\mathbf{F}_{_0}\mathbf{F}_{_0}^\prime\right]\right)\right]\\
&=&\mathbbm{E}\left[\text{tr}\left(\boldsymbol\Omega\left(\mathbf{F}_{_T}-\mathbf{F}_{_0}\right)\left(\mathbf{F}_{_T}-\mathbf{F}_{_0}\right)^\prime\right)\right],
\end{eqnarray*}
where the only requirement is that $\mathbf{F}$ follows a martingale (not necessarily an It\^o process). Finally, for all swaps associated with $\boldsymbol\beta$ we have
\begin{equation*}
\mathbbm{E}\left[\lim_{\boldsymbol\Pi_N\rightarrow\boldsymbol\Pi}\sum_{\boldsymbol\Pi_{_N}}\boldsymbol\gamma^\prime\left(\mathrm e^{\mathbf{\hat{x}}_i}-\mathbf{1}\right)\right]=\mathbbm{E}\left[\boldsymbol\gamma^\prime\left(\mathrm e^{\mathbf{x}_{_T}-\mathbf{x}_{_0}}-\mathbf{1}\right)\right]=0.
\end{equation*}
Therefore, if $\mathbf{z}=\left(\mathbf{F},\mathbf{x}\right)^\prime$, the necessary condition \eqref{eq:maincondition} is sufficient for all martingales. Note we can assume that $\boldsymbol\Omega$ is a symmetric matrix because $\text{tr}\left(\boldsymbol\Omega\mathbf{\hat{F}}\mathbf{\hat{F}}^\prime\right)$ is a quadratic form.\qed

\subsection{Proof of Theorem 3}\label{app:theorem3}

With the fair-value process of a DI swap contract is defined as
\begin{equation*}
V^\phi_t:=\mathbbm{E}_t\left[\sum_{\boldsymbol\Pi_{_N}}\phi\left(\mathbf{\hat{z}}\right)\right]-v^\phi_{_0},
\end{equation*}
the increments of this process along the partition $\boldsymbol\Pi_{_N}$ are given by
\begin{eqnarray*}
\hat{V}^\phi_i&=&V^\phi_{t_i}-V^\phi_{t_{i-1}}=\mathbbm{E}_{t_i}\left[\sum_{\boldsymbol\Pi_{_N}}\phi\left(\mathbf{\hat{z}}\right)\right]-\mathbbm{E}_{t_{i-1}}\left[\sum_{\boldsymbol\Pi_{_N}}\phi\left(\mathbf{\hat{z}}\right)\right]\nonumber\\
&=&\sum_{\tilde{i}=1}^{i}\phi\left(\mathbf{\hat{z}}_{\tilde{i}}\right)+\mathbbm{E}_{t_i}\left[\sum_{\tilde{i}=i+1}^N\phi\left(\mathbf{\hat{z}}_{\tilde{i}}\right)\right]-\sum_{\tilde{i}=1}^{i-1}\phi\left(\mathbf{\hat{z}}_{\tilde{i}}\right)-\mathbbm{E}_{t_{i-1}}\left[\sum_{\tilde{i}=i}^N\phi\left(\mathbf{\hat{z}}_{\tilde{i}}\right)\right]\\
&=&\phi\left(\mathbf{\hat{z}}_i\right)+\mathbbm{E}_{t_i}\left[\phi\left(\mathbf{z}_{_T}-\mathbf{z}_{t_i}\right)\right]-\mathbbm{E}_{t_{i-1}}\left[\phi\left(\mathbf{z}_{_T}-\mathbf{z}_{t_{i-1}}\right)\right]\nonumber\\
&=&\phi\left(\mathbf{\hat{z}}_i\right)+\hat{v}^\phi_i,
\end{eqnarray*}
where $\hat{v}^\phi_i=v^\phi_{t_i}-v^\phi_{t_{i-1}}$ and $v^\phi_t=\mathbbm{E}_t\left[\phi\left(\mathbf{z}_{_T}-\mathbf{z}_t\right)\right]$. Combining the above with Theorem 2 yields
\begin{eqnarray*}
\hat{v}^\phi_i&=&\mathbbm{E}_{t_i}\left[\boldsymbol\alpha^\prime\left(\mathbf{F}_{_T}-\mathbf{F}_{t_i}\right)+\text{tr}\left(\boldsymbol\Omega\left(\mathbf{F}_{_T}-\mathbf{F}_{t_i}\right)\left(\mathbf{F}_{_T}-\mathbf{F}_{t_i}\right)^\prime\right)+\boldsymbol\beta^\prime\left(\mathrm e^{\mathbf{x}_{_T}-\mathbf{x}_{t_i}}-\mathbf{1}\right)+\boldsymbol\gamma^\prime\left(\mathbf{x}_{_T}-\mathbf{x}_{t_i}\right)\right]\\
&&-\mathbbm{E}_{t_{i-1}}\left[\boldsymbol\alpha^\prime\left(\mathbf{F}_{_T}-\mathbf{F}_{t_{i-1}}\right)+\text{tr}\left(\boldsymbol\Omega\left(\mathbf{F}_{_T}-\mathbf{F}_{t_{i-1}}\right)\left(\mathbf{F}_{_T}-\mathbf{F}_{t_{i-1}}\right)^\prime\right)+\boldsymbol\beta^\prime\left(\mathrm e^{\mathbf{x}_{_T}-\mathbf{x}_{t_{i-1}}}-\mathbf{1}\right)\right.\\
&&\left.+\boldsymbol\gamma^\prime\left(\mathbf{x}_{_T}-\mathbf{x}_{t_{i-1}}\right)\right]\\
&=&\mathbbm{E}_{t_i}\left[\text{tr}\left(\boldsymbol\Omega\mathbf{F}_{_T}\mathbf{F}_{_T}^\prime\right)+\boldsymbol\gamma^\prime\mathbf{x}_{_T}\right]-\text{tr}\left(\boldsymbol\Omega\mathbf{F}_{t_i}\mathbf{F}_{t_i}^\prime\right)-\boldsymbol\gamma^\prime\mathbf{x}_{t_i}\\
&&-\mathbbm{E}_{t_{i-1}}\left[\text{tr}\left(\boldsymbol\Omega\mathbf{F}_{_T}\mathbf{F}_{_T}^\prime\right)+\boldsymbol\gamma^\prime\mathbf{x}_{_T}\right]+\text{tr}\left(\boldsymbol\Omega\mathbf{F}_{t_{i-1}}\mathbf{F}_{t_{i-1}}^\prime\right)+\boldsymbol\gamma^\prime\mathbf{x}_{t_{i-1}}\\
&=&\text{tr}\left(\boldsymbol\Omega\boldsymbol{\hat{\Sigma}}_i\right)+\boldsymbol\gamma^\prime\mathbf{\hat{X}}_i-\text{tr}\left(\boldsymbol\Omega\mathbf{F}_{t_i}\mathbf{F}_{t_i}^\prime\right)+\text{tr}\left(\boldsymbol\Omega\mathbf{F}_{t_{i-1}}\mathbf{F}_{t_{i-1}}^\prime\right)-\boldsymbol\gamma^\prime\mathbf{\hat{x}}_i,
\end{eqnarray*}
where $\boldsymbol{\hat{\Sigma}}_i=\boldsymbol{\Sigma}_{t_i}-\boldsymbol{\Sigma}_{t_{i-1}}$ with $\boldsymbol{\Sigma}_t=\mathbbm{E}_t\left[\mathbf{F}_{_T}\mathbf{F}_{_T}^\prime\right]$ and $\mathbf{\hat{X}}_i=\mathbf{X}_{t_i}-\mathbf{X}_{t_{i-1}}$ with $\mathbf{X}_t=\mathbbm{E}_t\left[\mathbf{x}_{_T}\right]$.
Thus
\begin{eqnarray*}
\hat{V}^\phi_i&=&\boldsymbol\alpha^\prime\mathbf{\hat{F}}_i+\text{tr}\left(\boldsymbol\Omega\left(\mathbf{F}_{t_i}-\mathbf{F}_{t_{i-1}}\right)\left(\mathbf{F}_{t_i}-\mathbf{F}_{t_{i-1}}\right)^\prime\right)+\boldsymbol\beta^\prime\left(\mathrm e^{\mathbf{\hat{x}}_i}-\mathbf{1}\right)+\boldsymbol\gamma^\prime\mathbf{\hat{x}}_i+\hat{v}^\phi_i\\
&=&\boldsymbol\alpha^\prime\mathbf{\hat{F}}_i+\text{tr}\left(\boldsymbol\Omega\left[\boldsymbol{\hat{\Sigma}}_i-2\mathbf{F}_{t_{i-1}}\mathbf{\hat{F}}_i^\prime\right]\right)+\boldsymbol\beta^\prime\left(\mathrm e^{\mathbf{\hat{x}}_i}-\mathbf{1}\right)+\boldsymbol\gamma^\prime\mathbf{\hat{X}}_i
\end{eqnarray*}
are the increments of the fair-value process for a \gls{di} swap on $\mathbf{z}=\left(\mathbf{F},\mathbf{x}\right)^\prime$.\qed

\subsection{Proof of Corollary}

The fair-value swap rate is
\begin{eqnarray*}
v^\phi_{_0}&=&\mathbbm{E}\left[\phi\left(\mathbf{z}_{_T}-\mathbf{z}_{_0}\right)\right]\\
&=&\mathbbm{E}\left[\boldsymbol\alpha^\prime\left(\mathbf{F}_{_T}-\mathbf{F}_{_0}\right)+\text{tr}\left(\boldsymbol\Omega\left(\mathbf{F}_{_T}-\mathbf{F}_{_0}\right)\left(\mathbf{F}_{_T}-\mathbf{F}_{_0}\right)^\prime\right)+\boldsymbol\beta^\prime\left(\mathrm e^{\mathbf{x}_{_T}-\mathbf{x}_{_0}}-\mathbf{1}\right)+\boldsymbol\gamma^\prime\left(\mathbf{x}_{_T}-\mathbf{x}_{_0}\right)\right]\\
&=&\mathbbm{E}\left[\text{tr}\left(\boldsymbol\Omega\left(\mathbf{F}_{_T}-\mathbf{F}_{_0}\right)\left(\mathbf{F}_{_T}-\mathbf{F}_{_0}\right)^\prime\right)+\boldsymbol\gamma^\prime\left(\mathbf{x}_{_T}-\mathbf{x}_{_0}\right)\right]\\
&=&\mathbbm{E}\left[\text{tr}\left(\boldsymbol\Omega\left[\mathbf{F}_{_T}\mathbf{F}_{_T}^\prime-\mathbf{F}_{_0}\mathbf{F}_{_0}^\prime\right]\right)+\boldsymbol\gamma^\prime\left(\mathbf{x}_{_T}-\mathbf{x}_{_0}\right)\right]\\
&=&\text{tr}\left(\boldsymbol\Omega\left[\boldsymbol\Sigma_{_0}-\mathbf{F}_{_0}\mathbf{F}_{_0}^\prime\right]\right)+\boldsymbol\gamma^\prime\left(\mathbf{X}_{_0}-\mathbf{x}_{_0}\right).\qed
\end{eqnarray*}

\subsection{Proof of Theorem 4}\label{app:theorem4}
Starting with
\begin{equation*}
\boldsymbol\Sigma_{_0}-\mathbf{F}_{_0}\mathbf{F}_{_0}^\prime=\left[
\begin{array}{ccc}
X_{_0}^{(2)}-X_{_0}X_{_0}&\ldots&X_{_0}^{(n)}-X_{_0}X_{_0}^{(n-1)}\\
\vdots&\ddots&\vdots\\
X_{_0}^{(n)}-X_{_0}X_{_0}^{(n-1)}&\ldots&X_{_0}^{(2n-2)}-X_{_0}^{(n-1)}X_{_0}^{(n-1)}
\end{array}\right],
\end{equation*}
for some $n\ge 2$, we use Theorem 3 as follows: 
\begin{eqnarray*}
v^\phi_{_0}&=&\mathbbm{E}\left[\phi\left(\mathbf{z}_{_T}-\mathbf{z}_{_0}\right)\right]=\text{tr}\left(\boldsymbol\Omega^{(n)}\left[\boldsymbol\Sigma_{_0}-\mathbf{F}_{_0}\mathbf{F}_{_0}^\prime\right]\right)\\
&=&\sum_{i=1}^{n-1}\omega_i^{(n)}\left(X_{_0}^{(i+1)}-X_{_0}X_{_0}^{(i)}\right)\\
&=&\omega_{n-1}^{(n)}X_{_0}^{(n)}+\sum_{i=2}^{n-1}\left(\omega_{i-1}^{(n)}-\omega_i^{(n)}X_{_0}\right)X_{_0}^{(i)}-\omega_1^{(n)}X_{_0}^2\\
&=&X_{_0}^{(n)}+\sum_{i=2}^{n-1}\tbinom{n}{i}\left(-X_{_0}\right)^{n-i}X_{_0}^{(i)}+(1-n)\left(-X_{_0}\right)^n\\
&=&\sum_{i=1}^n\tbinom{n}{i}\left(-X_{_0}\right)^{n-i}X_{_0}^{(i)}+\left(-X_{_0}\right)^n\\
&=&\mathbbm{E}\left[\sum_{i=0}^n\tbinom{n}{i}\left(-X_{_0}\right)^{n-i}x_{_T}^i\right]
=\mathbbm{E}\left[\left(x_{_T}-X_{_0}\right)^n\right]=v^{(n)}_{_0},
\end{eqnarray*}
where we have used $\omega_{n-1}^{(n)}=1$ and $\omega_1^{(n)}=\left(-X_{_0}\right)^{n-2}(n-1)$ in the third line.\qed

\end{appendix}
\end{document}